\documentclass[aps,prd,floats,floatfix, twocolumn,amssymb,amsmath,
superscriptaddress,nofootinbib,showpacs,longbibliography]{revtex4-2}

\usepackage[T1]{fontenc}
\usepackage[utf8]{inputenc}
\usepackage{txfonts} 
\usepackage[normalem]{ulem} 

\usepackage[usenames,dvipsnames]{xcolor}
\usepackage{graphicx}
\usepackage{tikz}
\usetikzlibrary{arrows.meta,positioning}
\usepackage[linktocpage,breaklinks]{hyperref}
\usepackage[capitalize]{cleveref}
\hypersetup{colorlinks=true,citecolor=NavyBlue,
linkcolor=NavyBlue,urlcolor=NavyBlue}

\usepackage{multirow}
\usepackage{array}[=2016-10-06] 
\usepackage{booktabs}
\DeclareMathAlphabet{\pazocal}{OMS}{zplm}{m}{n}
\usepackage{microtype}
\usepackage{subfigure}

\usepackage{journals}
\usepackage{upgreek}
\usepackage{stmaryrd} 
\newcommand{\nn}{\nonumber\\}
\newcommand{\dd}{{\rm d}}
\newcommand{\pa}{\partial}
\newcommand{\celephais}{\texttt{Celepha\"{\i}s}~}

\newcommand{\uiuc}{\affiliation{Department of Physics and Illinois Center for Advanced Studies of the Universe,\\University of Illinois Urbana-Champaign, Urbana, Illinois 61801, USA}}
\newcommand{\aei}{\affiliation{Max Planck Institute for Gravitational Physics (Albert Einstein Institute), D-14476 Potsdam, Germany}}

\begin{document}
\title{\texorpdfstring{\celephais}{Celephais}: efficient spectral initial data code for precessing compact binaries}

\date{\today}

\author{Hao-Jui Kuan}
\email{hjkuan@illinois.edu}
\uiuc \aei

\begin{abstract}
Large numerical-relativity surveys require compact-binary initial data that are both spectrally accurate and inexpensive to construct, including for systems with unequal masses and misaligned spins.
We present \celephais, a compact-object initial-data code built on the \texttt{Kadath} spectral library, that constructs binary-neutron-star and black-hole--neutron-star initial data without imposing equatorial symmetry.
The method exploits the sparse structure of the globally coupled multi-domain Jacobian and the approximate parity separation of fields.
The assembled matrix is factored with \texttt{MUMPS} and reused as a refreshed right preconditioner in a Jacobian-free Newton--Krylov iteration, thereby avoiding dense storage.
An adaptive $hp$--refinement scheme then concentrates resolution where the spectral tails are not yet resolved.
For a mass-ratio-$20$ black-hole--neutron-star benchmark, the adaptive schemes recover the uniform-grid constraint accuracy with about three times fewer unknowns.
We also extend the post-Newtonian force-balance estimate to arbitrary spin orientations and use it to initialise eccentricity reduction.
Validation comprises binding-energy comparisons with post-Newtonian sequences, a precessing binary-neutron-star eccentricity-reduction test, and a full evolution whose waveform-reconstructed precession axis follows a post-Newtonian simple-precession model.
These results establish an efficient route to spectrally resolved binary-neutron-star and black-hole--neutron-star initial data with arbitrary spin orientations.
\end{abstract}
\maketitle

\section{Introduction}
Gravitational-wave observations now probe compact-binary mergers throughout the strong-field regime.
As the observation catalogue grows, higher-signal-to-noise events and systems in less explored regions of parameter space are expected \cite{KAGRA:2021vkt,Kiendrebeogo:2023hzf}.
Accurate waveform models will require suites of numerical-relativity (NR) simulations spanning unequal masses, large and misaligned spins, eccentricity, and matter effects.

Each NR simulation begins with a solution of the Einstein constraint equations.
Residual constraint violations and orbital artefacts in these initial data propagate into the evolution and can limit waveform accuracy.
Multi-domain spectral methods are attractive because smooth solutions converge exponentially with spectral order \cite{Bonazzola:1998ge,Pfeiffer:2002wt,Grandclement:2007sb}.
The \texttt{LORENE} family established this approach for binary-neutron-star (BNS) and black-hole--neutron-star (BHNS) quasiequilibria \cite{LORENE,Gourgoulhon:2000nn,Grandclement:2001ed}.
The \texttt{Kadath} library subsequently provided a general spectral elliptic infrastructure \cite{Grandclement:2007sb,Grandclement:2009ju}, on which \texttt{FUKA} built public solvers for unequal-mass, aligned-spin compact binaries \cite{Papenfort:2021hod}.
Other spectral implementations that extend the physical or numerical scope include \texttt{SGRID} and \texttt{Elliptica}, which support generic spins and large mass ratios \cite{Dietrich:2015pxa,Tichy:2019ouu,Rashti:2021ihv}, while the \texttt{SpEC}/\texttt{Spells} framework uses a distinct multidomain infrastructure \cite{Pfeiffer:2002wt,Foucart:2008qt,Tacik:2015tja}.
Hyperbolic relaxation in \texttt{NRPyElliptic} \cite{Assumpcao:2021fhq} and the task-based discontinuous-Galerkin solver in \texttt{SpECTRE} \cite{Vu:2021coj} provide alternative routes to scalable elliptic solves.

Broad parameter surveys expose a tension between the robustness of a globally coupled spectral solve and the cost of resolving every domain uniformly.
A Newton step for a multi-domain spectral discretisation couples the elliptic equations and the multi-domain interface conditions through a global Jacobian.
Direct factorisation would requires $\mathcal{O}(N_u^2)$ storage for $N_u$ unknowns, even when their Newton Jacobians are mostly structurally zero, whereas matrix-free Newton--Krylov methods require an effective preconditioner \cite{Pfeiffer:2002wt,Vincent:2019qpd,Vu:2021coj}.
Rather than relying on a prescribed preconditioner, \celephais explicitly constructs the Jacobian for its strong-form elliptic system while assembling only the nonzero entries.
Like \texttt{FUKA}, it is a separate application code built on \texttt{Kadath}, but it extends the framework to arbitrary spin orientations and fully leverages the sparsity of the Jacobian and the parity properties of the fields.
Combined with adaptive $hp$--refinement, \celephais enables efficient construction of BNS and BHNS with substantially lower memory demand.
The resulting sparse matrix is factorised by \texttt{MUMPS} library, whose LU factors are then reused as a right preconditioner in the following Jacobian-free Newton--Krylov solve.
This way, we have the full Jacobian of the first Newton step as preconditioner to capture accurately the complicated couplings, and have the memory efficiency of a Jacobian-free solve for the subsequent steps.
In the event that the preconditioner becomes stale, the Jacobian is reassembled and refactored.
For a representative precessing BNS at production resolution, the first Jacobian assembly takes under $10$~minutes and the complete job uses under $17$~GB on a MacBook Pro with an Apple M4 Max processor.

Efficient linear algebra addresses only one part of the computational cost.
When underresolved spectral structure is confined to particular domains or coordinate directions, uniform refinement introduces unnecessary unknowns elsewhere.
In a high-mass-ratio binary, for example, the markedly unequal sizes of the two objects produce a broad region that is underresolved only in the radial direction.
Increasing a single uniform resolution parameter then adds coefficients to every domain and coordinate direction, including those whose spectral tails are already well resolved.
We implement an anisotropic $hp$--refinement algorithm that independently assesses the spectral tails of each domain and coordinate direction.
The adaptive mesh refinement (AMR) scheme especially helps to reach a target constraint accuracy with fewer unknowns for high-mass ratio binaries.

The remainder of the paper is organised as follows.
\Cref{sec:xcts} fixes notation by reviewing the extended conformal thin-sandwich (XCTS) formalism.
We then develop the dependency filter and batched Jacobian assembler in \cref{sec:filtering}.
\Cref{sec:parity,sec:jfnk_mumps} describe the parity-separated, refreshed-preconditioner Newton--Krylov solve and provide a benchmark of its performance.
\Cref{sec:amr} introduces the $hp$--refinement scheme.
We assess the implementation through a handful of examinations with increasing complexity in \cref{sec:assessment}.
\Cref{sec:conclusion} summarises the results and outstanding limitations.
Unless stated otherwise, we set $c=G=M_\odot=1$.

\section{Field equations in the XCTS formalism}
\label{sec:xcts}
To fix the notation used throughout the paper, we briefly recap the XCTS formalism, following closely the presentation of \texttt{FUKA} \cite{Papenfort:2021hod}.
We use the standard 3+1 decomposition \cite{Gourgoulhon:2007ue},
\begin{align}
  \dd s^2 = g_{\mu\nu}\dd x^\mu \dd x^\nu = -\alpha^2 \dd t^2
  + \gamma_{ij} \left(\dd x^i+\beta^i \dd t\right)
  \left(\dd x^j+\beta^j \dd t\right),
\end{align}
where $\alpha$ is the lapse, $\beta^i$ the shift, and $\gamma_{ij}$ the spatial metric on slices $\Sigma_t$ with future-pointing unit normal $n^\mu$.
Initial data satisfy the Hamiltonian and momentum constraints,
\begin{align}
  R + K^2 - K_{ij}K^{ij} &= 16\pi E, \label{eq:ham}\\
  D_j K^{j}{}_{i} - D_i K &= 8\pi J_i,
\end{align}
where $K_{ij}$ is the extrinsic curvature, $K=\gamma^{ij}K_{ij}$, and $D_i$ is the covariant derivative compatible with $\gamma_{ij}$.
The matter sources are defined as $E:=n^{\mu}n^{\nu}T_{\mu\nu}$ and $J_i:=-n_{\mu}\gamma_{\nu i}T^{\mu\nu}$.

The four constraints do not fix all twelve components of $\left(\gamma_{ij},K_{ij}\right)$.
The XCTS decomposition \cite{York:1998hy,Pfeiffer:2002iy} therefore separates freely specifiable data from variables determined by elliptic equations.
We write the conformal metric and trace-free curvature as
\begin{align}
  \tilde \gamma_{ij}&=\Psi^{-4} \gamma_{ij},\\
  \hat A_{ij} &= \Psi^{2} \left(K_{ij} - \tfrac{1}{3}K\gamma_{ij}\right)\,.
\end{align}
XCTS closes the four constraints by using the conformal trace-free metric evolution equation to relate $\hat A_{ij}$ algebraically to the lapse, shift, and freely specified conformal-metric velocity $\tilde u_{ij}:=\pa_t\tilde\gamma_{ij}$.
Specifying $\pa_tK$ turns the evolution equation for $K$ into a fifth elliptic equation for the lapse.
Thus, the system determines $\left(\Psi,\,\alpha\Psi,\,\beta^i\right)$ from $\left(\tilde\gamma_{ij},\,\tilde u_{ij},\,K,\,\pa_tK\right)$ and the matter sources.

For quasiequilibrium data, we set $\tilde u_{ij}=\pa_tK=0$, assuming that these freely specified fields are instantaneously stationary in the corotating frame associated with the approximate helical symmetry introduced below \cite{Cook:2001wi,Bonazzola:2003dm}.
We additionally adopt conformal flatness, $\tilde \gamma_{ij}=f_{ij}$ with $f_{ij}$ the flat metric, and maximal slicing, $K=0$.
Under these assumptions, the XCTS system reduces to the Isenberg--Wilson--Mathews form \cite{Isenberg:2007zg,Wilson:1995uh,Wilson:1996ty},
\begin{align}
  \tilde D^2\Psi &= -\tfrac{1}{8}\Psi^{-7}\hat A_{ij}\hat A^{ij}-2\pi\Psi^5 E, \label{eq:xcts:psi}\\
  \tilde D^2(\alpha\Psi) &= \tfrac{7}{8}\alpha\Psi^{-7}\hat A_{ij}\hat A^{ij}+2\pi\alpha\Psi^5(E+2S), \label{eq:xcts:lapse}\\
  \tilde D^2\beta^i &= -\tfrac{1}{3}\tilde D^i\tilde D_j\beta^j + 2\hat A^{ij}\tilde D_j(\alpha\Psi^{-6}) + 16\pi\alpha\Psi^4 J^i, \label{eq:xcts:shift}
\end{align}
where $\tilde D_i$ is the covariant derivative compatible with $\tilde\gamma_{ij}$ ($=f_{ij}$) and $\tilde D^2:=\tilde D_i \,\tilde D^i$.
The trace-free curvature follows from the shift through the conformal longitudinal operator $\tilde{\mathbb{L}}$,
\begin{align}
  \hat A^{ij} &= \frac{\Psi^6}{2\alpha}(\tilde{\mathbb{L}}\beta)^{ij},\\
  (\tilde{\mathbb{L}}v)^{ij} &:= \tilde D^i v^j+\tilde D^j v^i-\tfrac{2}{3}\tilde\gamma^{ij}\tilde D_k v^k,
\end{align}
and $S:=\gamma_{ij}S^{ij}$ is the trace of the spatial stress.

When the orbital separation changes on a timescale much longer than one orbit, the data may be constructed with an approximate helical Killing vector $\xi^\mu$, timelike inside the light cylinder \cite{Shibata:2004qz,Uryu:2005vv}.
In coordinates adapted to the symmetry, the time-evolution vector can be written as \cite{Bonazzola:1997gc,Tichy:2003zg}
\begin{align}
  \xi^\mu = \alpha n^\mu
  + \beta^\mu \,,
\end{align}
where the shift is split into an inertial piece $\beta_0^i$ and an analytic comoving background shift \cite{Ossokine:2015yla,Papenfort:2021hod},
\begin{align}
  \beta^i := \beta_0^i + \beta_{\rm co}^i \quad \text{with}\quad
  \beta^i_{\rm co} = \Omega\,\pa^i_\varphi \left(\boldsymbol{x}_c\right)
  + \frac{\dot a}{a} \left(x^i-x_c^i\right) + v_z \hat z^i\,.
  \label{eq:xcts:betacor}
\end{align}
Here $\Omega$ is the orbital angular velocity, $\pa^i_\varphi(\boldsymbol{x}_c)$ is the flat-space rotational vector about the binary centre $\boldsymbol{x}_c$, and $a$ is the coordinate separation.
Together, $\Omega$ and the radial approach speed $\dot a=\dd a/\dd t$ control the orbital eccentricity \cite{Moldenhauer:2014yaa,Buonanno:2010yk}.
The constant boost $v_z\hat z^i$ controls the ADM linear momentum normal to the orbital plane \cite{Ossokine:2015yla}, as described in \cref{sec:xcts:diag}.

For the conformally flat construction used here, the analytic shift in \cref{eq:xcts:betacor} keeps the compact objects stationary on the grid without changing the gravitational elliptic operators.
To see the latter aspect, the rotation and boost are flat-space Killing fields, while the radial term is a homothetic dilation.
These affine fields contribute neither to the shift Laplacian nor to the trace-free flat longitudinal operator: $(\tilde{\mathbb{L}}\beta_{\rm co})^{ij}=0$.
We therefore solve \cref{eq:xcts:shift} for $\beta_0^i$, reconstruct $\beta^i=\beta_0^i+\beta_{\rm co}^i$, and impose asymptotic flatness only on the regular variables,
\begin{align}
  \lim_{r\to\infty}\alpha  = 1\,, \quad
  \lim_{r\to\infty}\Psi    = 1\,, \quad
  \lim_{r\to\infty}\beta_0^i = 0 \,.
\end{align}
The analytic comoving terms instead encode the chosen orbital frame and need not decay at infinity.

\subsection{Matter sources and hydrostatic equilibrium}
\label{sec:xcts:matter}
We model neutron-star matter as a perfect fluid, $T^{\mu\nu}=\rho h\,u^\mu u^\nu+pg^{\mu\nu}$, with specific enthalpy $h$, rest-mass density $\rho$, specific internal energy $\epsilon$, pressure $p$, and four-velocity $u^\mu$.
The sources entering \eqref{eq:xcts:psi}--\eqref{eq:xcts:shift} are
\begin{align}
  E &= \rho h W^2 - p, \\
  S &= 3p+(E+p)U^2, \\
  J^i &= \rho h W^2 U^i,
\end{align}
where $U^i$ is the fluid velocity relative to the normal observer, $U^2:=\gamma_{ij}U^iU^j$, and the Lorentz factor is $W^2=(1-U^2)^{-1}$.
Because $\rho$ decreases steeply at the stellar surface, its direct spectral representation can develop Gibbs oscillations.
Following \cite{Tichy:2009yr,Tichy:2011gw,Tichy:2012rp,Moldenhauer:2014yaa,Dietrich:2015pxa,Papenfort:2021hod,Rashti:2021ihv}, we multiply the gravitational residuals by $p/\rho$.
This rescaling replaces explicit factors of $\rho$ in the source terms by smooth equation-of-state functions of $h$ and improves the surface representation.

Hydrostatic equilibrium further requires $\nabla_\mu T^{\mu\nu}=0$ and rest-mass conservation $\nabla_\mu(\rho u^\mu)=0$.
On each spatial slice, we define the projected enthalpy current
$\hat u_i:=h\gamma_i{}^\mu u_\mu=hWU_i$ and the fluid velocity in the corotating frame
$V^i:=\alpha U^i-\xi^i$.
For an isentropic fluid, the relativistic Euler equation can be written under helical symmetry as
\begin{align}
  D_i\!\left(\frac{h\alpha}{W}+\hat u_j V^j\right)+V^j\!\left(D_j\hat u_i-D_i\hat u_j\right)=0. \label{eq:xcts:euler}
\end{align}
For corotation, $V^i=0$ and the second term vanishes.
For irrotational flow, we introduce a velocity potential $\phi$ through
$\hat u_i=D_i\phi$ \cite{Shibata:1998um,Gourgoulhon:1998dr,Bonazzola:1998yq}.
The corotating velocity then becomes
\begin{align}
  V^i=\alpha U^i-\xi^i=\frac{\alpha}{\Psi^4hW}f^{ij}\tilde D_j\phi-\xi^i
\end{align}
and the antisymmetrised derivative of $\hat u_i$ vanishes.
Thus, \cref{eq:xcts:euler} admits an exact first integral for both corotation and irrotation.
Defining $H:=\ln h$, rest-mass conservation can be expressed as
\begin{align}
  \Psi^6 W V^i\tilde D_i H+\frac{\dd H}{\dd\ln\rho}\,\tilde D_i(\Psi^6 W V^i)=0\,, \label{eq:xcts:phieq}
\end{align}
which gives an elliptic equation for $\phi$ through $\tilde D_i V^i$ and avoids explicit dependence on the rest-mass density.

To prescribe neutron-star spin, we use the constant-rotational-velocity construction developed and applied in \cite{Tichy:2011gw,Tichy:2012rp,Foucart:2008qt,Foucart:2010eq,Foucart:2012vn,Tsatsin:2013jca,Kawaguchi:2015bwa,Tacik:2015tja,Dietrich:2015pxa,Dietrich:2017xqb,Kyutoku:2020xka,Rashti:2021ihv} and add a rotational component to the enthalpy current,
\begin{align}
  \hat u_i = D_i\phi + \Psi^4 f_{ij}s^j, \qquad s^i=\omega\,\xi^i_{\rm NS},
\end{align}
where $\omega$ parametrises the magnitude of the uniform rotation.
We allow $\xi^i_{\rm NS}$ to have arbitrary orientation and, within the helical-symmetry approximation, treat the spin contribution as constant along the neutron-star centre worldline \cite{Tichy:2011gw,Tichy:2012rp,Dietrich:2015pxa}.
Neglecting its spin-curl terms in \cref{eq:xcts:euler} gives the \emph{approximate} Bernoulli relation
\begin{align}
  \frac{h\alpha}{W}+D_i\phi\,V^i \simeq C_{\rm B} \,,
  \label{eq:xcts:firstint}
\end{align}
where $C_{\rm B}$ is constant on each star.
The elliptic field and matter equations do not determine the orbital angular velocity $\Omega$ or the position $\boldsymbol{x}_c$ of the rotation axis in \cref{eq:xcts:betacor}.
For a BNS force-balance solve, two stellar-centre conditions determine $\Omega$ and the component of $\boldsymbol{x}_c$ along the line of centres \cite{Gourgoulhon:2000nn,Dietrich:2015pxa,Papenfort:2021hod}.
For a BHNS, the single stellar condition determines $\Omega$, while the ADM-momentum conditions determine $\boldsymbol{x}_c$ \cite{Papenfort:2021hod}.
In either case, local force balance is imposed by requiring the enthalpy to be extremal along the line of centres at each neutron-star centre $\boldsymbol{x}_{{\rm NS}}$,
\begin{align}
  \left.\pa_x H\right|_{\boldsymbol{x}_{{\rm NS}}}=0\,.
\end{align}
Evaluating the $x$ derivative of \cref{eq:xcts:firstint} at each extremum balances the gravitational, orbital, and internal-flow terms along the line of centres.
The thermodynamic variables are closed separately by an equation of state (EOS) relating $\rho$, $p$, and $h$.

\subsection{Axisymmetric isolated-star XCTS system}
\label{sec:xcts:isolated_star}
The binary Newton solve requires an initial guess for each neutron star with the targeted EOS and spin.
We obtain this guess from an auxiliary isolated-star problem and use the resulting fields in the superposed binary data.
Because the isolated configuration is axisymmetric, the problem reduces to two dimensions.
With $\tilde{\beta}=\beta r\sin\theta$, the corresponding XCTS equations are \cite{Bonazzola:1993zz,Shibata:2007zzb,Gourgoulhon:2010ju}
\begin{align}
  \triangle_2 (\alpha \Psi)
  &= 2\pi \left( E + 2 S \right) \alpha \Psi^5
  + \frac{7 A^2}{4} \alpha \Psi^5\,,\\
  \triangle_2 \Psi
  &= - 2\pi E \Psi^5 - \frac{A^2}{4} \Psi^5 \,,\\
  \tilde{\triangle}_3 \tilde{\beta}
  &= \frac{16\pi \alpha J_\varphi}{r \sin\theta}
  + r\sin\theta\, (\pa\beta)\, [\pa(\ln\alpha-6\ln\Psi)] \,,
\end{align}
where $\triangle_2$ is the scalar flat-space Laplacian in axisymmetry and $\tilde{\triangle}_3$ is the $\varphi$--component of the vector Laplacian,
\begin{align}
  \triangle_2 &:=
  (\pa_r)^2
  + \frac{2}{r}\pa_r
  + \frac{1}{r^2}(\pa_\theta)^2
  + \frac{1}{r^2 \tan\theta}\pa_\theta\,,\\
  \tilde{\triangle}_3 &:=
  \triangle_2 - \frac{1}{r^2 \sin^2\theta}\,.
\end{align}
Here
\begin{align}
  A^2:=\frac{1}{2} K_{ij}K^{ij}=\frac{r^2\sin^2\theta}{4\alpha^2}
  \left( \pa\beta \right)^2 \,,
\end{align}
and we use the notation
\begin{align}
  \pa\alpha\,\pa\beta=\frac{\pa\alpha}{\pa r}\frac{\pa\beta}{\pa r}
  +\frac{1}{r^2}\frac{\pa\alpha}{\pa \theta}\frac{\pa\beta}{\pa \theta}\quad \text{and}\quad
  (\pa\alpha)^2=\pa\alpha\,\pa\alpha\,.
\end{align}

\subsection{Boundary conditions at compact-object surfaces}
For black holes, we excise a coordinate 2-sphere $S_{\rm BH}$ and impose quasiequilibrium inner boundary conditions \cite{Gourgoulhon:2001ec,Cook:2004kt,Jaramillo:2004uc,Grandclement:2022wif}.
We require $S_{\rm BH}$ to be a marginally outer trapped surface, i.e.\ a surface of vanishing outgoing null expansion.
In the conformally flat XCTS variables, this requirement gives the Robin condition
\begin{align}
  \tilde s^i\tilde D_i\Psi\big|_{S_{\rm BH}} &= -\frac{\Psi}{4}\tilde D^i\tilde s_i-\frac{1}{4}\Psi^{-3}\hat A_{ij}\tilde s^i\tilde s^j\,,
\end{align}
where $\tilde s^i$ is the conformal unit normal to $S_{\rm BH}$.
Keeping the excision surface at a fixed coordinate location imposes
\begin{align}
  \beta^i\big|_{S_{\rm BH}} &= \alpha\Psi^{-2}\tilde s^i+\Omega_{\rm BH}\xi^i_{\rm BH}\,.
\end{align}
Here $\xi^i_{\rm BH}$ is tangent to the horizon and $\Omega_{\rm BH}$ sets its rotation, while the first term fixes the normal component of the shift.
On the spherical horizon, we choose the homogeneous Neumann gauge condition
\begin{align}
  \tilde s^i\tilde D_i(\alpha\Psi)\big|_{S_{\rm BH}} &= 0 \,.
\end{align}
On the other hand, neutron stars require no excision.
Their surfaces are defined by $H=0$ and, at each surface, \cref{eq:xcts:phieq} reduces regularly to
\begin{align}
  V^i\tilde D_i H=0 \,.
\end{align}

\subsection{Global and quasi-local diagnostics}
\label{sec:xcts:diag}
We use global charges as indicators of quasiequilibrium and momentum balance, and quasi-local charges to fix the compact-object parameters.
Under conformal flatness, the ADM and Komar masses at spatial infinity are \cite{Arnowitt:1960zzc,DeWitt:1967yk,OMurchadha:1974pq,York:1978gql}
\begin{align}
  M_{\rm ADM} &= -\frac{1}{2\pi}\oint_{S_\infty}\tilde D^i\Psi\,\dd S_i,\\
  M_{\rm K} &= \frac{1}{4\pi}\oint_{S_\infty}\tilde D^i\alpha\,\dd S_i.
\end{align}
where $\dd S_i$ is the outward conformal surface element.
Approximate helical symmetry requires $M_{\rm ADM}=M_{\rm K}$, and the dimensionless discrepancy
\begin{align}
  \epsilon_{\rm vir} =
  \frac{|M_{\rm K}-M_{\rm ADM}|}{M_{\rm K}+M_{\rm ADM}} \,,
  \label{eq:virial}
\end{align}
is monitored as a virial error.
The total angular and ADM linear momenta are \cite{Arnowitt:1962hi,Regge:1974zd,Gourgoulhon:2012ffd}
\begin{align}
  J_{\rm tot} &= \frac{1}{8\pi}\oint_{S_\infty}\hat A^{ij}\xi_i\,\dd S_j,\\
  P^i_{\rm ADM} &= \frac{1}{8\pi}\oint_{S_\infty}\hat A^{ij}\,\dd S_j .
\end{align}
The binding energy is $E_b=M_{\rm ADM}-M_\infty$, where $M_\infty$ is the sum of the individual masses at infinite separation.

We achieve $P_x=P_y=0$ by adjusting the centre-of-mass position in the rigid co-orbital velocity.
Because rotation about $\hat{\boldsymbol z}$ generates no velocity along that axis, this adjustment cannot control $P_z$ in generic tilted-spin configurations.
We therefore solve for a uniform boost $v_z\hat z^i$ and impose the corresponding momentum constraint at infinity, which is the out-of-plane part of the boost-based centre-of-mass control of \cite{Ossokine:2015yla}.

The remaining compact-object parameters in \cref{eq:xcts:psi,eq:xcts:lapse,eq:xcts:shift} are fixed by quasi-local mass and spin constraints.
For each object, an approximate rotational vector $\xi^i_{\rm (NS,BH)}$ is centred on the object and aligned with the requested spin axis; in the aligned case, it reduces to $\pa^i_\varphi(\boldsymbol{x}_c)$.
For a black hole, the spin and irreducible mass on the excision surface $S_{\rm BH}$ are
\begin{align}
  \mathcal{S} &= \frac{1}{8\pi}\oint_{S_{\rm BH}}\hat A_{ij}\xi^i_{\rm BH}\,\dd S^j, \label{eq:spin}\\
  M_{\rm irr}^2 &= \frac{1}{16\pi}\oint_{S_{\rm BH}}\Psi^4\,\dd S,
\end{align}
where $\dd S^j=\tilde s^j\dd S$ and $\dd S$ is the conformal area element, giving the Christodoulou mass and dimensionless spin
\begin{align}
  M_{\rm CH}^2 = M_{\rm irr}^2+\frac{\mathcal{S}^2}{4M_{\rm irr}^2}, \qquad \chi=\frac{\mathcal{S}}{M_{\rm CH}^2}.
\end{align}
For a neutron star, the same spin integral \eqref{eq:spin}, evaluated on a coordinate sphere $S_{\rm NS}$ enclosing all matter, gives the quasi-local spin $\mathcal{S}_{\rm QL}$ \cite{Tacik:2015tja,Tichy:2019ouu}.
A spinning BNS solve imposes $\chi_{\rm NS}=\mathcal{S}_{\rm QL}/M_{{\rm ADM},{\rm NS}}^2$ for each star.
The baryonic mass is the volume integral
\begin{align}
  M_{\rm b} = \int_{V_{\rm NS}} W\rho\,\Psi^6\,\dd V,
\end{align}
where $\dd V$ is the conformal volume element.

\section{Sparse Jacobian construction and MUMPS-preconditioned Newton--Krylov solve}
\label{sec:method}
\begin{figure*}
  \centering
  \resizebox{2\columnwidth}{!}{%
  \begin{tikzpicture}[
    every node/.style={font=\tiny},
    lbl/.style={font=\tiny, align=center},
    shell/.style={draw, thick},
    ncf/.style={draw, thick, dash dot},
    adapt/.style={draw, dashed},
    lead/.style={-{Stealth[length=1.4mm]}, thin, gray},
    hmove/.style={{Stealth[length=1.4mm]}-{Stealth[length=1.4mm]}, thick}
  ]
    \draw[dashed] (0,0) circle (3.9);
    \draw[shell] (0,0) circle (3.3);
    \draw[ncf] (0,0) circle (2.6);
    \draw[thin] (-1.37,2.21) .. controls (-1.67,1.70) and (-1.55,0.85) .. (-1.295,0.445);
    \draw[thin] (-1.37,2.21) .. controls (-0.76,1.84) and (-0.07,0.78) .. (-0.055,0);
    \draw[thin] ( 1.37,2.21) .. controls ( 0.76,1.84) and ( 0.07,0.78) .. ( 0.055,0);
    \draw[thin] ( 1.37,2.21) .. controls ( 1.67,1.70) and ( 1.55,0.85) .. (1.27,0.43);
    \draw[thin] (-1.37,-2.21) .. controls (-1.67,-1.70) and (-1.55,-0.85) .. (-1.295,-0.445);
    \draw[thin] (-1.37,-2.21) .. controls (-0.76,-1.84) and (-0.07,-0.78) .. (-0.055,0);
    \draw[thin] ( 1.37,-2.21) .. controls ( 0.76,-1.84) and ( 0.07,-0.78) .. ( 0.055,0);
    \draw[thin] ( 1.37,-2.21) .. controls ( 1.67,-1.70) and ( 1.55,-0.85) .. (1.27,-0.43);
    \node[font=\scriptsize] at (-2.00,0.90) {1};
    \node[font=\scriptsize] at (-0.98,1.25) {2};
    \node[font=\scriptsize] at (0,1.74) {3};
    \node[font=\scriptsize] at (0.98,1.25) {4};
    \node[font=\scriptsize] at (2.06,0.90) {5};
    \fill (0,0) circle (0.045);
    \fill[Gray!25] (-0.85,0) circle (0.5);
    \draw (-0.85,0) circle (0.5);
    \draw[adapt] (-0.85,0) circle (0.63);
    \fill[Gray!25] (0.85,0) circle (0.25);
    \draw (0.85,0) circle (0.25);
    \draw[adapt] (0.85,0) circle (0.6);
    \draw[-{Stealth[length=1.2mm]}, thin] (0.85,0) --
      node[pos=1,right,font=\scriptsize] {$R_\star$} (1.067,0.125);
    \draw[-{Stealth[length=1.2mm]}, thin] (0.85,0) --
      node[pos=1,above,font=\scriptsize] {$R_{\rm bisph}$} (0.85,0.6);
    \draw[hmove] (0,2.6) -- (0,3.3);
    \foreach \r in {2.72,2.84,2.95,3.07,3.18,3.3}
      \fill ({\r*0.574},{\r*0.819}) circle (0.024);
    \node[lbl] (Lcomp) at (0,4.55) {compactified domain $\to$ spatial infinity};
    \draw[lead] (Lcomp.south) -- (0,3.9);
    \node[lbl] (Lh) at (-3.3,3.25) {$h$-move:\\insert shell};
    \draw[lead] (Lh.east) -- (-0.05,2.95);
    \node[lbl] (Lp) at (3.5,2.95) {$p$-move:\\add points};
    \draw[lead] (Lp.west) -- (1.55,2.25);
    \node[lbl] (Lext) at (5.0,0.7) {exterior shell\\($h$ or $p$ in $r$)};
    \draw[lead] (Lext.west) -- (2.95,0.55);
    \node[lbl] (Lncf) at (-4.9,1.6) {non-conforming\\interface};
    \draw[lead] (Lncf.east) -- (-2.13,1.49);
    \node[lbl] (Lcf) at (-5.0,-0.5) {conforming\\interface};
    \draw[lead] (Lcf.east) -- (-2.54,-0.54);
    \node[lbl] (Lstar) at (-4.6,-2.5) {neutron star\\surface-fitted, one $(n_\theta,n_\phi)$};
    \draw[lead] (Lstar.north) -- (-0.95,-0.5);
    \node[lbl] (Lbi) at (4.6,-2.3) {five bispheric domains\\$\phi$-block, shared grid};
    \draw[lead] (Lbi.north) -- (0.0,-0.95);
  \end{tikzpicture}}
  \caption{Binary multi-domain decomposition and admissible refinement moves (schematic, not to scale).
  Two surface-fitted stars are connected by bispheric domains and surrounded by one spherical shell and a compactified outer domain.
  The field--domain blocks are the dependency units used in \cref{sec:filtering}; the arrows indicate the admissible $h$ and $p$ refinements discussed in \cref{sec:amr}.
  Thin internal curves delimit the five bispheric domains, numbered in the upper half of the schematic.
  On the smaller object, $R_\star$ extends from its centre to the object surface and $R_{\rm bisph}$ from the same centre to the surrounding dashed boundary.
  The large dash-dotted circle marks the non-conforming interface; solid circles denote conforming interfaces.
  An interactive 3D version is available at \url{https://hao-jui.github.io/Celephais/SpaceDecomposition.html}.}
  \label{fig:amr_domains}
\end{figure*}
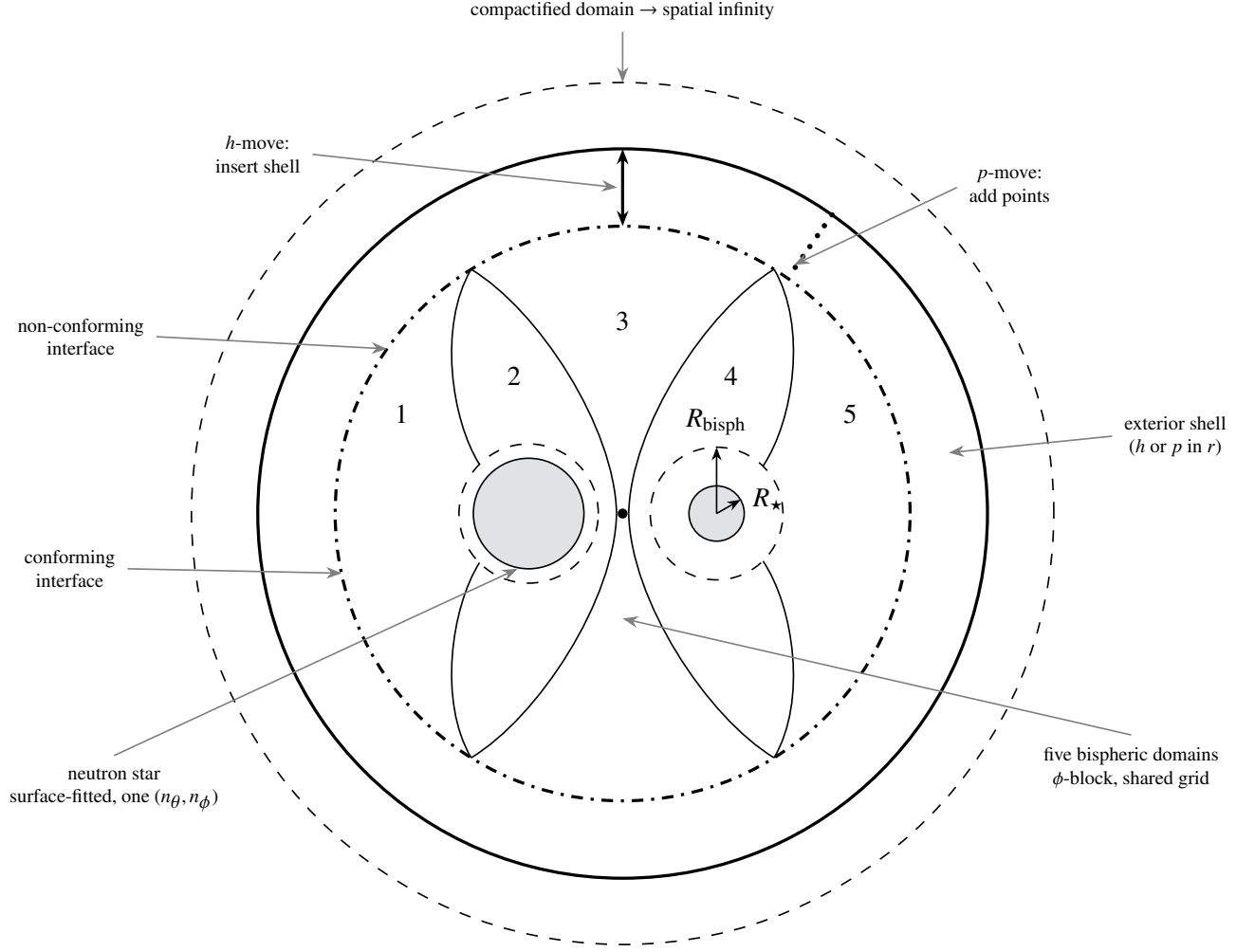

\texttt{KADATH} discretises the XCTS system on the multi-domain space shown in \cref{fig:amr_domains}, using surface-adapted coordinates to track neutron-star boundaries \cite{Uryu:1999uu,Gourgoulhon:2000nn,Ansorg:2003br}.
At Newton step $k$, the state $u_k$ contains the spectral fields, stellar-surface shapes, and scalar parameters, while the residual $r_k$ collects the volume equations, boundary and matching conditions, and global constraints.
With the code's internal sign convention, the correction satisfies
\begin{align}
    J_k\,\delta u_k = r_k\,,
    \qquad
    u_{k+1}=u_k-\delta u_k\,.
\end{align}
This correction is globally coupled, but its Jacobian is structurally sparse.
We exploit this structure by estalishing an economical equation-to-Jacobian machienry through a dependency graph identifying potentially nonzero derivatives, batched automatic differentiation evaluating them, separating the factorisation into invariant sectors under a $y$--reflection parity consideration, and a periodically refreshed sparse factorisation preconditions the matrix-free Newton--Krylov solve.

\subsection{Operator dependency graph and structural filtering}
\label{sec:filtering}
\begin{figure}[t]
  \centering
  \resizebox{\columnwidth}{!}{%
  \begin{tikzpicture}[
    every node/.style={font=\small},
    title/.style={font=\large\bfseries, align=center},
    op/.style={draw, rounded corners=1pt, align=center, minimum width=1.35cm, minimum height=0.55cm},
    defn/.style={draw, dashed, rounded corners=1pt, align=center, minimum width=1.35cm, minimum height=0.55cm},
    leaf/.style={draw, align=center, minimum width=1.35cm, minimum height=0.55cm},
    bucket/.style={draw, rounded corners=2pt, align=left, inner sep=6pt},
    result/.style={draw, rounded corners=1pt, align=center, inner sep=6pt},
    arr/.style={-{Stealth[length=1.5mm]}, thin}
  ]
  \node[title] (tree_title) at (0,0) {recursive scan of leaves};
  \node[op] (eq) at (0,-0.9) {$E_i$};
  \node[defn] (d2) at (-1.25,-1.8) {$\mathsf{Q}_2$};
  \node[op] (der) at (1.25,-1.8) {$\pa_a$};
  \node[defn] (d1) at (-1.25,-2.7) {$\mathsf{Q}_1$};
  \node[leaf] (qa) at (-2.05,-3.6) {$q_A(d_1)$};
  \node[leaf] (const) at (-0.45,-3.6) {const.};
  \node[leaf] (qb) at (1.25,-2.7) {$q_B(d_2)$};
  \draw[arr] (eq) -- (d2);
  \draw[arr] (eq) -- (der);
  \draw[arr] (d2) -- (d1);
  \draw[arr] (d1) -- (qa);
  \draw[arr] (d1) -- (const);
  \draw[arr] (der) -- (qb);

  \node[title] (bucket_title) at (5.25,0) {support bucket};
  \node[bucket] (sets) at (5.25,-1.75) {$\mathcal{V}(\mathsf{Q}_1)=\{q_A(d_1)\}$\\
    $\mathcal{V}(\mathsf{Q}_2)=\{q_A(d_1)\}$\\[0.15em]
    $\mathcal{V}(E_i)=\{q_A(d_1),q_B(d_2)\}$};
  \draw[arr] (der.east) -- (sets.west);

  \node[result] (filter) at (5.25,-3.45) {for varied $q_j(d_j)$,\\
  evaluate $E_i$ only if\\
  $q_j(d_j)\in\mathcal{V}(E_i)$};
  \draw[arr] (sets) -- (filter);
  \end{tikzpicture}}
  \caption{Dependency filter for a residual operator $E_i$.
  The symbolic expression is traversed through intermediate definitions to its unknown field--domain leaves, producing the support $\mathcal{V}(E_i)$.
  When $q_j(d_j)$ is varied, $E_i$ is evaluated only if $q_j(d_j)\in\mathcal{V}(E_i)$.}
  \label{fig:operator_dependency_scan}
\end{figure}
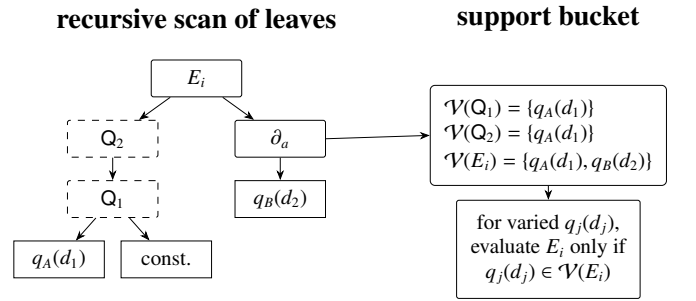

We assemble the sparse Jacobian by columns and determine its structural zeros before differentiating.
This is possible because \texttt{KADATH} represents each residual $E_i$ as a symbolic expression.
The scan illustrated in \cref{fig:operator_dependency_scan} follows the expression through its intermediate definitions $\mathsf{Q}_\lambda$ to the unknown field--domain pairs on which it can depend.
The resulting acyclic dependency graph defines supports $\mathcal{V}(\mathsf{Q}_\lambda)$ and $\mathcal{V}(E_i)$.

\begin{figure}
  \centering
  \includegraphics[width=\columnwidth]{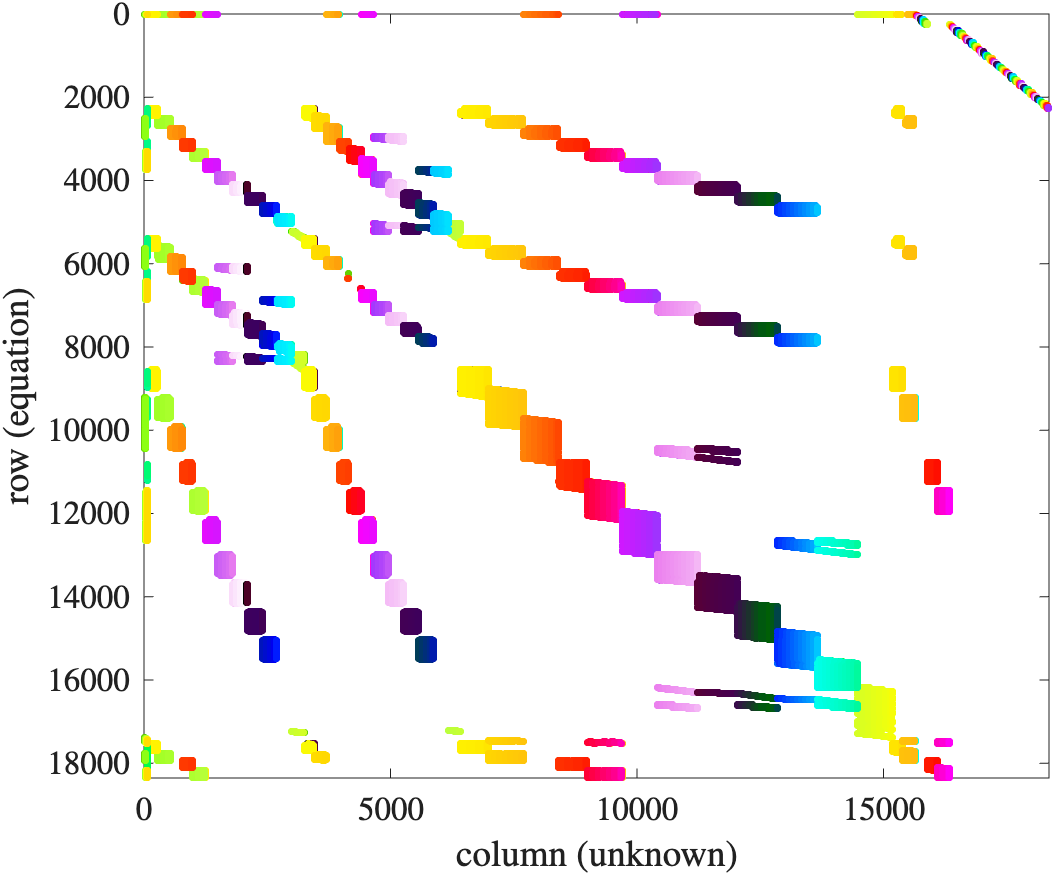}
  \caption{Sparse Jacobian for a low-resolution equal-mass BNS with one spinning star.
  Only $2.1\%$ of the matrix entries are nonzero, and colours identify columns grouped by the batched automatic-differentiation assembler.
  Batching reduces the number of residual traversals from $18\,355$ scalar sweeps to $2980$ grouped sweeps. }
  \label{fig:spy}
\end{figure}

When a coefficient of $q_j(d_j)$ is varied, only definitions and residual rows whose supports contain that field--domain pair are evaluated, and a structural zero is thusly excluded.
Forward-mode automatic differentiation evaluates the derivatives passing the dependency filter.
For a discrete residual $r=F(u)$, the $j$--th Jacobian column is the directional derivative of $F(u+\varepsilon\mathbf{e}_j)$ at $\varepsilon=0$, restricted to the residual rows that may contribute.
Because the first variation is linear in its seed, strip-mined forward mode \cite{Griewank:2008ed} propagates several compatible columns in one traversal, $\dot R = J\,\dot U = \big[\,J\mathbf{e}_{j_1},\dots,J\mathbf{e}_{j_{N_{\rm lane}}}\,\big]$ for $\dot U = \big[\,\mathbf{e}_{j_1},\dots,\mathbf{e}_{j_{N_{\rm lane}}}\,\big]$, so the primal operations are shared while each lane carries one Jacobian column.
\Cref{fig:spy} illustrates the resulting sparsity and reduction in residual traversals, using a representative low-resolution BNS as example.
Several strides of non-zero entries are visible, which are the representation of one equation is a domain.
We only batch the columns when they share the same field--domain support to reconcile with the structural-zero filter.

\subsection{Reflection parity and sector-separated preconditioning}
\label{sec:parity}
Before factorisation, the sparse preconditioner can be separated into two sectors when the binary admits a reflection symmetry.
We place the binary axis along $x$, take the orbital angular momentum along $z$, and restrict the spin axes to the $x$--$z$ plane.
For $(\boldsymbol{x}_c)_y=0$ and vanishing radial-infall parameter $\dot a=0$, the residual map is equivariant under reflection $y\to-y$ combined with reversal of velocity-type quantities; we denote this transformation by $\mathcal{P}_y$.
The symmetry applies independently of the mass ratio, spin magnitudes, and spin tilts within the $x$--$z$ plane.

Each spectral degree of freedom is assigned the product of its field parity and the parity of its azimuthal basis function under $\mathcal{P}_y$.
The same combined parity is assigned to the residual rows, thereby partitioning both unknowns and equations into even and odd sectors.
Denoting $F(u)$ as the discrete residual map, equivariance means $F(\mathcal{P}_y u)=\mathcal{P}_yF(u)$.
At an invariant state $\mathcal{P}_y u=u$, differentiating this relation gives a Jacobian that commutes with the parity operation.
When the unknowns and rows are ordered by this combined parity,
\begin{align}
  [J,\mathcal{P}_y]=0,
  \qquad
  J=
  \begin{pmatrix}
    J_{+} & 0 \\
    0 & J_{-}
  \end{pmatrix} .
\end{align}
Here $J_{+}$ and $J_{-}$ act on the even and odd sectors.
On the first sparse assembly, we check that cross-sector entries are consistent with roundoff and then remove them before factorising both sectors; neither sector is discarded.

The radial approach $\dot a\ne0$ used in eccentricity reduction (\cref{sec:ecc_red}) weakly breaks this symmetry.
In that case, the sector mask is used only for the quasi-Newton preconditioner, while the matrix-free action of $J_k$ retains the physical cross-sector coupling.
This approximation did not prevent convergence for the weakly noncircular configurations considered here.
It is worth cautioning that the mixed-parity components of solutions are retained, and are only truncated at the Jacobian level.
If the measured coupling is appreciable or the parity assignment is inconsistent, the unmodified preconditioner of \cref{sec:jfnk_mumps} is used.

\subsection{MUMPS-preconditioned Newton--Krylov solve}
\label{sec:jfnk_mumps}
We solve the Newton correction with unrestarted generalised minimal residual iteration (GMRES).
High-order spectral derivatives, boundary and matching conditions, and global integral rows have different numerical scales, which motivates preconditioning the matrix-free Krylov solve.
The measurements below assess only the preconditioned implementation and do not provide an unpreconditioned robustness comparison.

At a refresh step $k_f$, the filtered and batched assembler of \cref{sec:filtering} constructs the sparse Jacobian $J_{k_f}$.
When the parity mask is active, \texttt{MUMPS} factors the two sectors introduced above separately; otherwise it factors the full matrix, using a \texttt{METIS} ordering in either case \cite{1998SJSC...20..359K}.
We denote the matrix represented by these factors by $M_{k_f}:=J_{k_f}$; applying $M_{k_f}^{-1}$ requires sparse triangular solves, not an explicit inverse.
Between refreshes, the factors are reused in the right-preconditioned system
\begin{align}
    \delta u_k = M_{k_f}^{-1} y_k \,,
    \qquad
    J_k\, M_{k_f}^{-1}\, y_k = r_k \,.
\end{align}
Here $M_{k_f}^{-1}$ uses the latest stored factors, whereas $J_k$ is always evaluated matrix-free at the current Newton state, i.e., every Newton step uses the current Jacobian and the cached information is only the preconditioner.
Reusing the factorisation therefore changes the Krylov convergence rate but not the Newton equation being solved.

The Krylov tolerance is tied to the nonlinear residual through the Eisenstat--Walker choice-2 forcing term \cite{DemboEisenstatSteihaug1982,EisenstatWalker1996}.
In the production configuration used here, GMRES stops when
\begin{align}
  \left\|r_k-J_kM_{k_f}^{-1}y_k^{(m)}\right\|_2
  \leq \eta_k\left\|r_k\right\|_2,
  \qquad m\leq 48,
\end{align}
without restart before the cap.
The forcing term is bounded between $10^{-8}$ and $10^{-3}$ and tightens as the nonlinear residual contracts, avoiding oversolving early Newton steps.
Failure to meet the criterion is reported, and the resulting step is not counted as a converged linear solve.
The first Jacobian is always factored; thereafter, the factors are reused while nonlinear convergence remains rapid and are refreshed periodically or when the residual grows sharply.

\Cref{tab:benchmarks} reports the cost of constructing the first sparse preconditioner for a representative precessing BNS before AMR.
Increasing the MPI rank count accelerates Jacobian assembly but not as much for the sparse factorisation.
Actually, using many ranks for \texttt{MUMPS} can be rather inefficient, because the factorisation is a competition between computation and communication where the latter dominates at high rank counts.
Therefore, we intented to use one fourth of $n_{\rm p}$, rounded up to the least integer, for \text{MUMPS}.
At $N=15$, the factorisation is infeasible within the memory budget, and we need to use the out-of-core functionality of \texttt{MUMPS}.
The out-of-core action writes the partial factors to disk while factorising other factors and reads them back when needed for the triangular solves.
In general, the factorisation time will be slowed by the disk I/O than the in-core factorisation.
This headroom is negligible for Macbook Pro used here, but this is not a portable conclusion and depends on the I/O bandwidth of the machine.
The rapid growth in the number of unknowns with uniform order motivates the adaptive refinement introduced in \cref{sec:amr}.

\begin{table}
  \centering
  \footnotesize
  \setlength{\tabcolsep}{0pt}
  \renewcommand{\arraystretch}{1.15}
  \begin{tabular*}{\columnwidth}{@{\extracolsep{\fill}}r r r r r r@{}}
  \toprule
  \multicolumn{6}{@{}l}{\textit{Strong scaling at fixed resolution $N=11$}} \\
  \midrule
  $n_{\rm p}$ & Jacob. (s) & Wall time (s) & RSS (GB) & speedup & efficiency \\
  \midrule
  1 & 47.77 & 61.35 & 4.442 & $1.00\times$ & 100.0\% \\
  2 & 26.81 & 38.76 & 4.213 & $1.78\times$ & 89.1\% \\
  4 & 14.90 & 27.66 & 3.960 & $3.21\times$ & 80.2\% \\
  8 & 8.85  & 24.03 & 3.489 & $5.40\times$ & 67.5\% \\
  \midrule
  \multicolumn{6}{@{}l}{\textit{Resolution sweep at fixed rank count $n_{\rm p}=2$}} \\
  \midrule
  Res $N$ & DOF & Jacob. (s) & \texttt{MUMPS} (s) & RSS (GB) & OOC (s) \\
  \midrule
  9  & 41\,612  & 9.09   & 2.14  & 1.397  & 2.12  \\
  11 & 81\,035  & 26.81  & 11.68 & 4.213  & 11.69 \\
  13 & 120\,168 & 69.15  & 43.40 & 13.873 & 47.20 \\
  15 & 191\,200 & 163.53 & ---   & ---    & 150.08\\
  \bottomrule
  \end{tabular*}
  \caption{Sparse-Jacobian construction and \texttt{MUMPS} analysis plus factorisation for an equal-mass $1.35+1.35\,M_\odot$ precessing BNS with the DD2 EOS and coordinate separation $35\,M_\odot$.
  One star is nonspinning; the other has $\chi=0.3$ inclined by $81^\circ$ to the orbital angular momentum.
  Values are medians of three runs on a MacBook Pro with an Apple M4 Max processor.
  The strong-scaling block varies the MPI rank count at $N=11$, where speedup and efficiency are computed from Jacobian assembly.
  The resolution block uses two ranks, where ``---'' indicates that the factorisation was infeasible within the claimable memory (RAM).
  However, the use of out-of-core (OOC) memory enables the $N=15$ factorisation to complete.
  }
  \label{tab:benchmarks}
\end{table}

\section{Adaptive hp-refinement}
\label{sec:amr}
\begin{figure}
\centering
\resizebox{\columnwidth}{!}{%
\begin{tikzpicture}[
  every node/.style={font=\small},
  pt/.style={fill, circle, inner sep=0pt, minimum size=2.6pt},
  bar/.style={thick},
  ar/.style={-{Stealth[length=1.8mm]}}
]
  \node[anchor=east] at (-0.3,2.4) {$p$-move};
  \draw[bar] (0,2.4) -- (3,2.4);
  \draw (0,2.22)--(0,2.58); \draw (3,2.22)--(3,2.58);
  \foreach \x in {0.5,1.5,2.5} \node[pt] at (\x,2.4){};
  \draw[ar] (3.35,2.4) -- (3.95,2.4);
  \draw[bar] (4.25,2.4) -- (7.25,2.4);
  \draw (4.25,2.22)--(4.25,2.58); \draw (7.25,2.22)--(7.25,2.58);
  \foreach \x in {4.6,5.1,5.6,6.1,6.6,7.1} \node[pt] at (\x,2.4){};
  \node[anchor=east] at (-0.3,0.8) {$h$-move};
  \draw[bar] (0,0.8) -- (3,0.8);
  \draw (0,0.62)--(0,0.98); \draw (3,0.62)--(3,0.98);
  \foreach \x in {0.5,1.5,2.5} \node[pt] at (\x,0.8){};
  \draw[ar] (3.35,0.8) -- (3.95,0.8);
  \draw[bar] (4.25,0.8) -- (7.25,0.8);
  \draw (4.25,0.62)--(4.25,0.98); \draw (7.25,0.62)--(7.25,0.98);
  \draw[very thick] (5.75,0.55)--(5.75,1.05);
  \foreach \x in {4.6,5.0,5.4} \node[pt] at (\x,0.8){};
  \foreach \x in {6.1,6.5,6.9} \node[pt] at (\x,0.8){};
\end{tikzpicture}}
\caption{The two admissible radial refinement moves on one domain (schematic).
A $p$ move raises the order in place, whereas an $h$ move inserts a radial interface.
The polar and azimuthal directions admit only $p$ moves.}
\label{fig:amr_hp}
\end{figure}
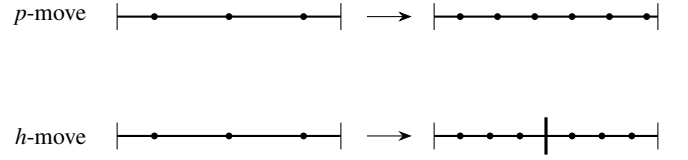

Sparse linear algebra reduces the cost at a fixed grid, but uniform resolution can allocate DOF inefficiently.
The required order depends on the binary parameters, and is not known \textit{a priori}.
We therefore use domain-local spectral tails to refine only unresolved domains and coordinate directions.
Following \cite{Renkhoff:2023nfw}, a $p$ move adds collocation points along one coordinate direction, an $h$ move subdivides a radial shell, and their combination is $hp$ refinement (\cref{fig:amr_hp}).
The domain decomposition constrains the admissible refinement moves, as summarised in \cref{tab:amr}.
The layout in \cref{fig:amr_domains} contains two surface-fitted stars and their surrounding shells, five connecting bispheric domains \cite{Ansorg:2005bp,Grandclement:2009ju}, exterior shells, and a compactified outer domain.
A $p$ move raises the order of an existing domain, whereas an $h$ move subdivides only the shellable stellar-side and exterior bands that do not include the bispheric domains.

Refinement must also preserve the tau-method interface conditions \cite{Grandclement:2007sb,Grandclement:2009ju}.
Equal angular orders are matched mode by mode.
When neighbouring orders differ, the fields are instead interpolated to common boundary points before their difference is projected into the tau rows.
These interfaces enforce that all domains belonging to one stellar core retain a common angular resolution, and that the five bispheric domains refine together and only in $\phi$.
The former preserves the single angular representation of the deformable stellar surface, while the latter preserves conformity within the bispheric block.

An $h$ move helps when sufficient radial space remains between the compact object, at radius $R_\star$, and the bispheric matching surface, at $R_{\rm bisph}$.
After inserting $n_{\rm shell}$ interfaces, this interval is divided into $n_{\rm shell}+1$ equal-width shells, preserving the domain geometry over the configurations considered here.

\subsection{Refinement indicators and marking strategy}
\label{sec:amr-indicators}

\begin{table}
  \centering
  \footnotesize
  \setlength{\tabcolsep}{4pt}
  \renewcommand{\arraystretch}{1.25}
  \begin{tabular}{@{}p{2.5cm}ccc p{3.45cm}@{}}
  \toprule
  \raggedright domain group & $r$ & $\theta$ & $\phi$ & \raggedright coupling rule \tabularnewline
  \midrule
  \raggedright nucleus + adapted pair & $p$ & $p$ & $p$ & \raggedright one $(n_\theta,n_\phi)$ for the stellar core \tabularnewline
  \raggedright stellar-side shells & $h{+}p$ & $p$ & $p$ & \raggedright shellable band before the bispheric match \tabularnewline
  \raggedright bispheric ($\times 5$) & --- & --- & $p$ & \raggedright refine as a block, $\phi$ only \tabularnewline
  \raggedright exterior shells & $h{+}p$ & $p$ & $p$ & \raggedright interfaces matched by the tau method \tabularnewline
  \raggedright compactified & $p$ & $p$ & $p$ & \raggedright outermost cell, reaches $\infty$ \tabularnewline
  \bottomrule
  \end{tabular}
  \caption{AMR moves allowed by the binary domain decomposition; ``---'' denotes a direction that cannot be refined.}
  \label{tab:amr}
\end{table}
For a sufficiently regular resolved field, spectral coefficients decay rapidly, so the highest retained modes provide a local estimate of truncation error \cite{Renkhoff:2023nfw}.
Let $c_\alpha$ denote the coefficients on one domain, with multi-index $\alpha=(\alpha_r,\alpha_\theta,\alpha_\phi)$.
Along coordinate direction $\ell$, we define the tail from the highest $w=2$ modes,
\begin{align}
  \mathcal{T}_\ell=\{\,\alpha:\alpha_\ell\ge n_\ell-w\,\},
  \qquad \ell\in\{r,\theta,\phi\}.
\end{align}
For each nonzero spectrum, the $L_2$ and $L_\infty$ tail ratios are \cite{PerssonPeraire:2006}
\begin{align}
  \eta_{2,\ell}
  &=\frac{\big(\sum_{\alpha\in\mathcal{T}_\ell}|c_\alpha|^2\big)^{1/2}}
          {\big(\sum_\alpha|c_\alpha|^2\big)^{1/2}},
  &
  \eta_{\infty,\ell}
  &=\frac{\max_{\alpha\in\mathcal{T}_\ell}|c_\alpha|}
          {\max_\alpha|c_\alpha|}.
\end{align}
For tensor fields, both norms include all components.

The default tolerances are $\tau_2=10^{-8}$ and $\tau_\infty=10^{-7}$ in every direction.
For solved field $q$, domain $d$, and direction $\ell$, the normalised demand is
\begin{align}
  \mathcal{D}_{d,\ell}:=\max_q\max\!\left(
  \frac{\eta^{(q)}_{2,d,\ell}}{\tau_2},
  \frac{\eta^{(q)}_{\infty,d,\ell}}{\tau_\infty}
  \right).
\end{align}
Pairs with $\mathcal{D}_{d,\ell}>1$ are marked for refinement, while the largest demand provides a global convergence diagnostic.
For the azimuthal Fourier basis, cosine and sine coefficients are combined into the physical amplitude of each mode before the tail is evaluated.

\begin{figure}
  \centering
  \includegraphics[width=\columnwidth]{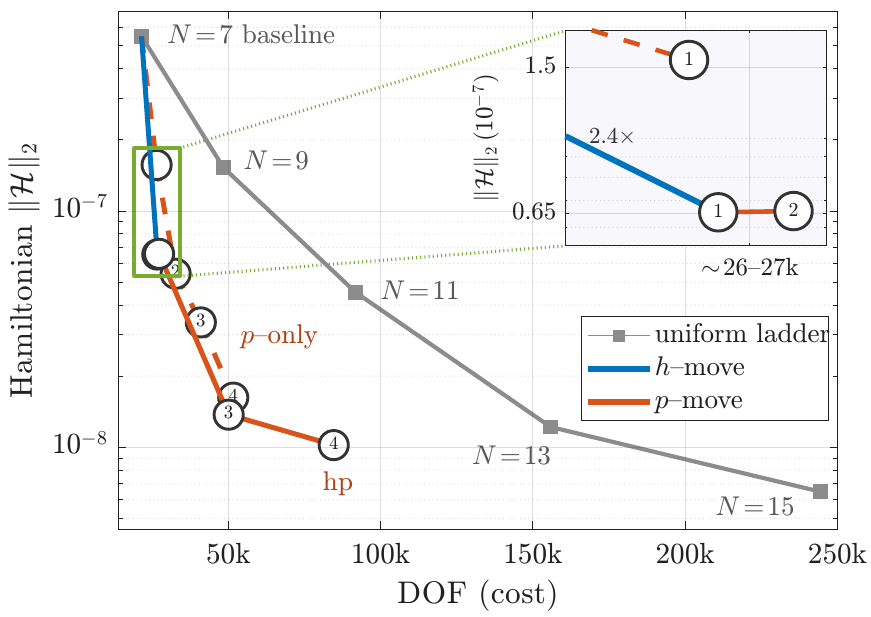}
  \caption{Cost--accuracy relation for a mass-ratio-$20$ BHNS with the DD2 EOS and coordinate separation $240\,M_\odot$.
  The Hamiltonian-constraint norm $\|\mathcal{H}\|_2$ is shown against the number of unknowns for uniform, $p$-only, and $hp$ refinement from a common $N=7$ baseline.
  A radial imbalance triggers one $h$ move, which inserts three shells at fixed order and reduces $\|\mathcal{H}\|_2$ by about a factor of $8$, to approximately $1/2.4$ of that obtained by the equal-cost first $p$ move.
  Subsequent $p$ moves reach the finest uniform-grid accuracy with about three times fewer unknowns.}
  \label{fig:amr_cost_q20}
\end{figure}

\begin{figure}
  \centering
  \begin{minipage}[t]{0.48\textwidth}
    \centering
    \includegraphics[width=\textwidth]{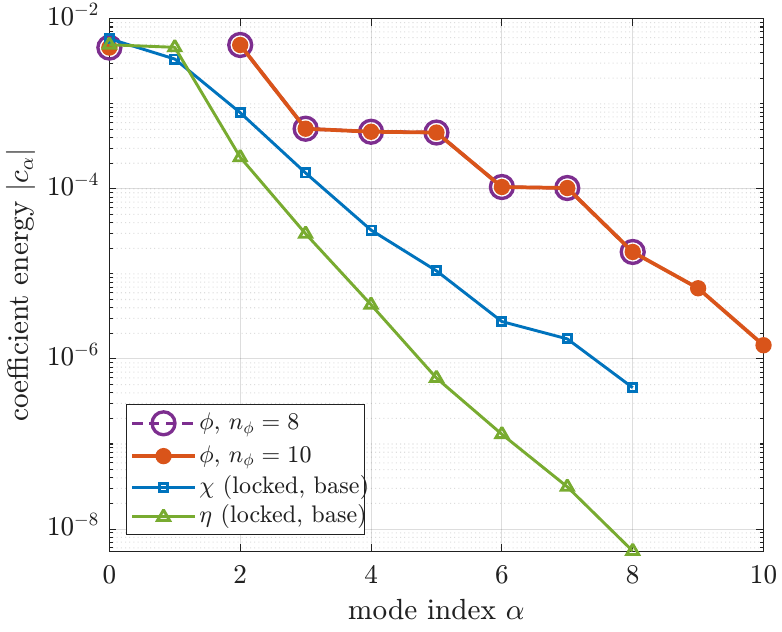}
  \end{minipage}
\hfill
  \begin{minipage}[t]{0.48\textwidth}
    \centering
    \includegraphics[width=\textwidth]{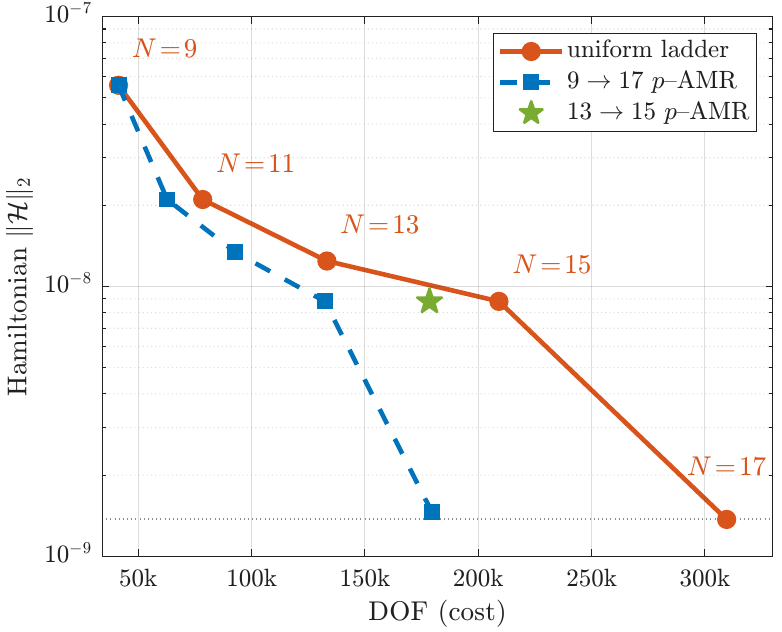}
  \end{minipage}
  \caption{Local and global effects of $p$ refinement.
  \textbf{Left}: directional spectral coefficients on a bispheric domain for the precessing BNS of \cref{tab:benchmarks}.
  The limiting azimuthal tail triggers $n_\phi=8\!\to\!10$, reducing the tail by about one order of magnitude while leaving the resolved modes unchanged.
  \textbf{Right}: Hamiltonian-constraint norm against unknown count for a nonspinning mass-ratio-$10$ BHNS with a $1.5\,M_\odot$ neutron star, the DD2 EOS, and coordinate separation $200\,M_\odot$.
  Anisotropic $p$ refinement reaches the uniform $N=15$ accuracy with about $1.6$ times fewer unknowns.}
  \label{fig:spectral_tail}
\end{figure}
\subsection{Execution policy}
\label{sec:amr-policy}
An $h$ move inserts one radial shell into each marked shellable band, whereas a $p$ move adds two points to a marked domain--direction pair.
To select between them, let $\mathcal{R}_k$ be the largest radial demand among the $h$--eligible domains at AMR cycle $k$, and let $\mathcal{A}_k$ be the largest polar or azimuthal demand over all domains.
When a shell budget remains and $\mathcal{A}_k<d_A\mathcal{R}_k$, the policy uses an $h$ move; otherwise it uses anisotropic $p$ refinement.
We set the angular-dominance threshold to $d_A=1$.

An accepted $h$ move must reduce the radial demand in every affected band by at least a factor of two before further subdivision is allowed.
If this test fails, the $h$ gate is closed and the next cycle uses a restricted $p$ fallback, avoiding a broad order increase immediately after the radial layout changes.
For either $p$ branch, candidates are considered from largest to smallest demand, as in a capped analogue of D{\"o}rfler marking \cite{Doerfler:1996}.
The estimated coefficient-count growth is limited to a factor $\Gamma_p=2$ for an ordinary $p$ move and $\Gamma_{p,h}=1.1$ after an $h$ move, with a per-axis order cap $n_{\max}=15$.
The cycles end when no pair remains marked or no admissible candidate fits these limits.

We illustrate how radial subdivision improves efficiency when the domain layout is imbalanced, using a BHNS with a mass ratio of 20 in \cref{fig:amr_cost_q20}.
Comparing with the uniform resolution sequence, both pure $p$ and $hp$ refinements help to reach the same level of accuracy, represented by the Hamiltonian-constraint residual, with less cost.
Noticeably, the first $h$ move yields an order of magnitude reduction in $\|\mathcal{H}\|_2$, the $L^2$ norm of the disagreement between the left- and right-hand sides of \Cref{eq:ham}.

On the other hand, \cref{fig:spectral_tail} demonstrates the benefit of $p$ move both locally and globally.
In the top panel, we zoom in to one of the bispheric domain for a precessing BNS, where we see the coefficient tail of $\phi$--direction is improved by more than an order of magnitude by locally increasing the $\phi$ resolution. 
The improvement in the global solution is exemplified by a BHNS with a mildly high mass ratio of 10.
We compare the uniform ladder from $N=9$ to 17.
Alongside, the $p$--refined solutions from the $N=9$ baseline always cost considerably less while keeping the same level of accuracy.
We also include a solution obtained by applying a single $p$ refinement to an
$N=13$ baseline. 
It indicates that the resulting accuracy is largely independent of the refinement path once the fields reach $N=15$ in the domain and spectral direction that limit convergence.

\section{Numerical assessment of quasiequilibrium quality}
\label{sec:assessment}
Spectral convergence of the elliptic solve is necessary but does not by itself establish quasiequilibrium quality.
The previous section tested spectral convergence for BHNS systems with $q:=m_2/m_1=20$ and $q=10$.
We now apply four increasingly dynamical checks: compatibility with \texttt{FUKA} data, PN binding-energy consistency, eccentricity reduction in a short evolution (\cref{sec:ecc_red}), and a full evolution from which we extract the waveform (\cref{sec:sim}) and reconstruct the precession axis (\cref{sec:orbital_axis}).

\subsection{\texttt{FUKA} compatibility and binding-energy consistency}
\label{sec:pn_assessment}
As a compatibility test, we initialise \celephais with the \texttt{FUKA} BNS data used in \cite{Kuan:2024jnw,Kuan:2025bzu} and show that they are still solutions to the new code.
This check is available only for the aligned-spin configurations \texttt{FUKA} supports \cite{Papenfort:2021hod}.
The precessing and tilted-spin data that motivate this work lie outside that overlap and admit no \texttt{FUKA} reference solution.
For those data, assessment instead relies on the eccentricity-reduction, evolution, and waveform-based consistency tests below (\cref{sec:ecc_red,sec:sim,sec:orbital_axis}).

\begin{figure}
  \centering
  \includegraphics[width=\columnwidth]{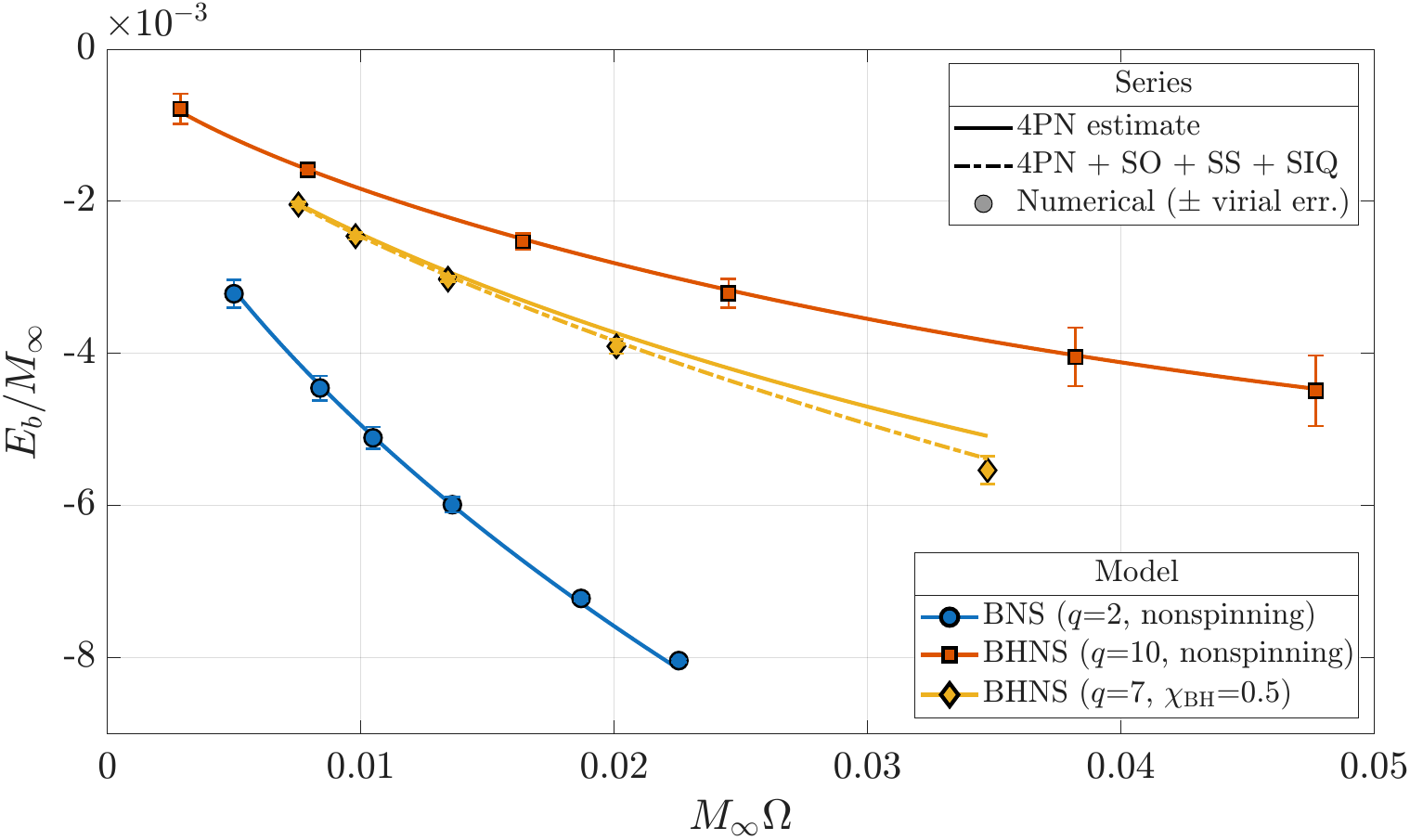}
  \caption{Binding energy $E_b/M_\infty$ against orbital frequency $M_\infty\Omega$ for three $N=13$ quasicircular sequences: a nonspinning BNS with $q=2$ and $M_\infty=3$; a nonspinning BHNS with $q=10$ and $M_\infty=16.5$; and a BHNS with $q=7$, $M_\infty=12$, and black-hole spin $\chi_{\rm BH}=0.5$.
  All neutron stars use the DD2 EOS.
  Symbols show the numerical binding energies, and the vertical bars show the virial discrepancy of \cref{eq:virial} as a scale diagnostic rather than a statistical uncertainty.
  Solid curves show the nonspinning 4PN estimate \cite{Blanchet:2013haa}. 
  For the spinning BHNS, the dashed curve also includes the spin--orbit, spin--spin, and spin-induced-quadrupole terms \cite{Kidder:1995zr,Bohe:2012mr,Papenfort:2021hod}.
  The numerical sequences follow the corresponding PN trends over the sampled frequencies.}
  \label{fig:pn}
\end{figure}

\Cref{fig:pn} then compares quasicircular \celephais binding-energy sequences with their PN estimates before any evolution-based eccentricity correction.
For the most demanding configurations tested, agreement requires either a uniform $N=13$ grid or an $N=11$ baseline followed by one AMR cycle.
We therefore use this as the minimum production-resolution criterion for the configurations considered here; its adequacy outside the tested set must be checked with the same diagnostics.

\subsection{Eccentricity reduction}
\label{sec:ecc_red}

Low-eccentricity data generally require iteration rather than a single quasicircular elliptic solve.
For binary black holes, one evolves trial data for several orbits, fits the residual oscillation in separation or orbital frequency, and updates the orbital frequency and radial approach rate \cite{Pfeiffer:2007yz}.
Precession complicates the fit because spin-driven modulations occur near twice the orbital frequency, whereas eccentricity appears near the orbital frequency \cite{Buonanno:2010yk}.
Orbital- and waveform-based estimators reduce, but do not remove, this gauge dependence \cite{Purrer:2012wy,Ramos-Buades:2018azo,Habib:2024soh}.
Nonetheless, the fitting formula used in \cite{Papenfort:2021hod} is adopted here.

For BNS and BHNS data, the analogous procedure varies the orbital angular velocity and an approach velocity within the hydrostationary problem, then calibrates them with short evolutions \cite{Kyutoku:2014yba,Kyutoku:2020lgg,Moldenhauer:2014yaa}.
In the \texttt{Kadath}/\texttt{FUKA} construction, $\Omega$ and $\dot a$ therefore serve as eccentricity-control parameters: PN estimates provide the initial values, but dynamical measurements determine the final correction \cite{Papenfort:2021hod}.
A spin-dependent initial estimate is particularly useful for precessing BNS because each calibration evolution is expensive.

\celephais evaluates algebraic PN estimates for $\Omega$ and $\dot a$ at the requested coordinate separation $a=|\boldsymbol{x}|:=|\boldsymbol{x}_1-\boldsymbol{x}_2|$, where $\boldsymbol{x}_{1,2}$ are the compact-object centres and $\hat{\boldsymbol r}:=\boldsymbol{x}/a$.
The approach speed enters \cref{eq:xcts:betacor} through the homothetic rate $\dot a/a$.
For aligned or antialigned spins, symmetry supplies the nonradial circularity conditions.
Instantaneous radial balance combines the nonspinning circular baseline with the leading spin--orbit and spin--spin accelerations from Kidder's PN equations of motion \cite{Kidder:1995zr},
\begin{align}
  0
  &= -a\Omega_0^2(a)
  + a_{\rm SO}^r(a,\Omega,\boldsymbol{S}_1,\boldsymbol{S}_2)
  + a_{\rm SS}^r(a,\boldsymbol{S}_1,\boldsymbol{S}_2)
  + a\Omega^2 \,.
\label{eq:balance}
\end{align}
Here $\Omega_0(a)$ is the nonspinning 3PN value \cite{Blanchet:2001id,Papenfort:2021hod}; $a_{\rm SO}^r$ and $a_{\rm SS}^r$ are the radial spin--orbit and spin--spin projections; and $\boldsymbol{S}_A:=m_A^2\boldsymbol{\chi}_A$ is the angular momentum of object $A$, with mass $m_A$ and dimensionless spin $\boldsymbol{\chi}_A=\chi_A\hat{\boldsymbol{s}}_A$.
The positive root of \cref{eq:balance} gives the initial estimate of $\Omega$.

The comoving shift also requires $\dot a$.
At the same separation, we begin with the nonspinning value $\dot a_0(a)$ and apply Kidder's leading aligned-spin circular-inspiral correction \cite{Kidder:1995zr},
\begin{align}
  \dot a(a,\boldsymbol{\chi}_1,\boldsymbol{\chi}_2)
  = \dot a_0(a)
  \left[ 1-\mathcal{C}_{\rm SO} \left(\frac{m}{a}\right)^{3/2}
  \right] \,, \label{eq:infall}
\end{align}
with
\begin{align}
  \dot a_0(a)
  &= -\frac{64}{5}\frac{m^3\eta}{a^3}
  \left[
  1+\frac{m}{a}\left(-\frac{1751}{336}-\frac{7}{4}\eta\right)
  \right] \,, \\
  \mathcal{C}_{\rm SO}
  &= \frac{7}{12}\sum_{A=1}^{2}\chi_A
  \left(\hat{\boldsymbol{L}}_N\cdot\hat{\boldsymbol{s}}_A\right)
  \left(19\frac{m_A^2}{m^2}+15\eta\right) \,,
\end{align}
where $m=m_1+m_2$, $\eta=m_1m_2/m^2$, and $\hat{\boldsymbol{L}}_N$ is the orbital-angular-momentum direction.

For precessing configurations, the spin accelerations must be evaluated as vectors before taking the radial projection.
With $\boldsymbol{a}_{\rm rel}:=\dd^2\boldsymbol{x}/\dd t^2$, the conservative relative acceleration at a trial angular velocity is
\begin{align}
\boldsymbol{a}_{\rm rel}
= -a\Omega_0^2(a)\,\hat{\boldsymbol{r}}
 + \boldsymbol{a}_{\rm SO}
   \left(a,\Omega,\hat{\boldsymbol{r}},
         \boldsymbol{S}_1,\boldsymbol{S}_2\right)
 + \boldsymbol{a}_{\rm SS}
   \left(a,\hat{\boldsymbol{r}},
         \boldsymbol{S}_1,\boldsymbol{S}_2\right) \,,
\end{align}
where $\boldsymbol{a}_{\rm SO}$ and $\boldsymbol{a}_{\rm SS}$ are the leading spin--orbit and spin--spin vectors from the same PN equations \cite{Kidder:1995zr}.
We then solve only the instantaneous radial balance condition for $\Omega$; the nonradial spin acceleration describes the precessional dynamics and is not set to zero.
For $\dot a$, the present implementation still uses \cref{eq:infall} with each spin projected onto $\hat{\boldsymbol L}_N$.
This aligned-spin approximation is a principal limitation of the precessing initial estimate.

\begin{figure}
  \centering
  \includegraphics[width=\columnwidth]{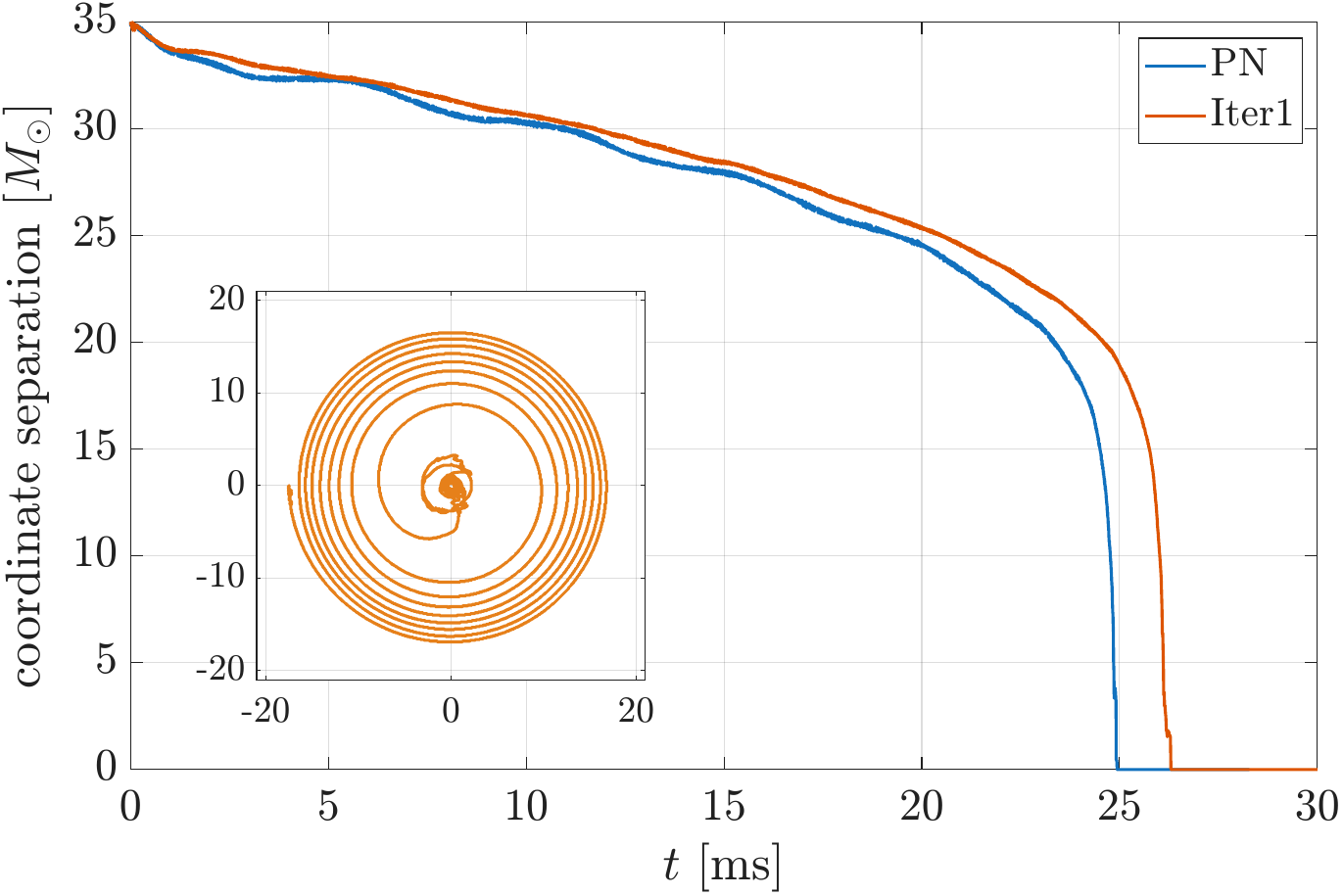}
  \caption{Eccentricity reduction for the same precessing BNS as in \cref{tab:benchmarks}.
  The coordinate separation $a(t)$ is shown for data initialised directly by the PN estimates of $\Omega$ and $\dot a$ [\cref{eq:balance,eq:infall}] and after one evolution-based correction (Iter1).
  The correction fits the component of $a(t)$ near the orbital frequency and updates both control parameters.
  The inset shows the projected trajectory of the spinning neutron star, which completes more than seven orbits before merger.}
  \label{fig:ecc}
\end{figure}

An evolution-based iteration handles the final calibration.
We use the procedure of \cite{Papenfort:2021hod}, fitting the separation oscillation near the orbital frequency and correcting $\Omega$ and $\dot a$.
\Cref{fig:ecc} shows the same precessing BNS as in \cref{tab:benchmarks}.
The PN-initialised data give a measured eccentricity below $10^{-2}$, and one evolution-based correction removes most of the visible orbital-frequency oscillation.
This single example demonstrates a useful starting estimate, not a guarantee that one iteration suffices throughout precessing parameter space.

\subsection{Dynamical evolution and waveform extraction}
\label{sec:sim}
A full dynamical evolution provides the most direct check that the data are usable.
We evolve the same precessing BNS as in \cref{tab:benchmarks} with the graphics-processor-accelerated \texttt{SACRAK} code \cite{Han:2026vnm}, developed from the \texttt{SACRA} \cite{Yamamoto:2008js}, \texttt{SACRA-MPI} \cite{Kiuchi:2017pte,Kiuchi:2019kzt,Kiuchi:2022ubj}, and \texttt{NANASI} \cite{Kiuchi:2025ksk}.
The data support a stable inspiral through to merger, and supply the waveform multipoles that \cref{sec:orbital_axis} uses to reconstruct the precession axis.
For the simulations reported here, we adopt the resolution with a finest spacing of $\Delta x=0.125M_\odot\simeq180$~m.

The waveform is derived from the Weyl scalar $\psi_4$, extracted at finite radius and decomposed into spin-weighted spherical-harmonic modes.
The leading perturbative correction of Nakano \cite{Nakano:2015pta} approximates each mode at future null infinity,
\begin{align}
  \left. r\psi_4^{\ell m}\right|_{r=\infty}
  &= \left(1-\frac{2M_\infty}{r}\right)\nn
  &\times\left[ r\psi_4^{\ell m}(t,r)
   - \frac{(\ell-1)(\ell+2)}{2r}\int \dd t\, r\psi_4^{\ell m}(t,r)\right] \,,
\end{align}
where $r$ is the areal radius, $M_\infty$ is the background mass, and we omit the contribution of background spin here.
The strain modes follow from $\psi_4^{\ell m}=\ddot h_{\ell m}$ by fixed-frequency double integration \cite{Reisswig:2010di}.
The low-frequency cutoff is set as $0.8|m|\Omega_{\rm orb}^{\rm ini}$ to suppress secular drift, where $\Omega_{\rm orb}^{\rm ini}$ is initial orbital angular frequency.
For a precessing binary, we express the waveform in a quadrupole-aligned (QA) frame.
If $h^{\rm QA}_{\ell m}$ denotes a co-precessing mode and the instantaneous radiation-axis proxy is described by $(\theta_{\rm L},\varphi_{\rm L})$, the inertial strain along a fixed line of sight can be written as~\cite{OShaughnessy:2011pmr,Kawaguchi:2017wrt}
\begin{align}
  h(t)
  &=
  \sum_{\ell,m}
  e^{-2i\varphi_{\rm L}(t)}
  {}_{-2}Y_{\ell m}\!\left[-\theta_{\rm L}(t),-\psi_{\rm L}(t)\right]
  h^{\rm QA}_{\ell m}(t), \nn
  \psi_{\rm L}(t)
  &=
  -\int^t \dot\varphi_{\rm L}(t')\cos\theta_{\rm L}(t')\,\dd t' ,
  \label{eq:qa_projection}
\end{align}
where the third Euler angle $\psi_{\rm L}$ enforces minimal rotation.
This convention separates precession of the radiation axis from an arbitrary rotation about it.

Modes are referred to retarded time $t_{\rm ret}=t-r_*$, where $r_*=r+2M_\infty\ln\!\left(r/2M_\infty-1\right)$ is the Schwarzschild tortoise coordinate of the extraction sphere.
Fixed-frequency integration uses a Tukey window that vanishes for $t_{\rm ret}<t_{\rm cut}$ and reaches unity after an interval $\Delta t$, suppressing the initial-data transient and the window turn-on.
We begin the analysis at $t_{\rm start}=t_{\rm cut}+\Delta t$, align the phase by imposing $\phi(t_{\rm start})=0$, and take $t_{\rm cut}=0$ and $\Delta t=200\,M_\infty\simeq2.8$~ms.
All quantities below use $t_{\rm ret}\ge t_{\rm start}$.

\subsection{Waveform reconstruction of the precession axis}
\label{sec:orbital_axis}
\begin{figure}[t]
  \centering
  \includegraphics[width=\columnwidth]{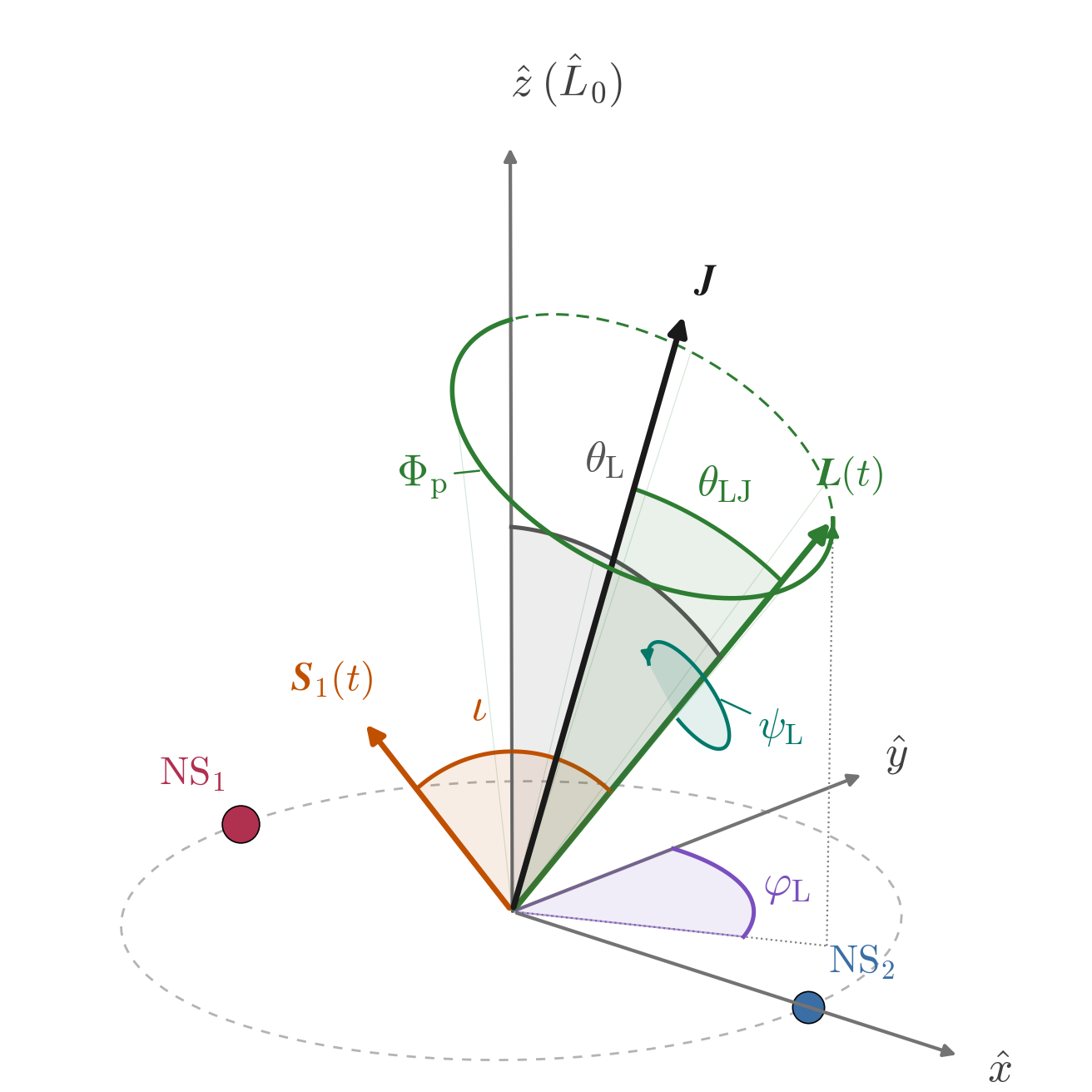}
  \caption{Angle conventions for the single-spin precessing frame.
  At the reference time the stars lie on the $\hat{\boldsymbol x}$ axis, $\boldsymbol L\parallel\hat{\boldsymbol z}$, and $\boldsymbol S_1$ has spin--orbit tilt $\iota$.
  In the simple-precession model, $\boldsymbol L(t)$ and $\boldsymbol S_1(t)$ rotate at fixed mutual tilt about the nearly conserved $\boldsymbol J=\boldsymbol L+\boldsymbol S_1$.
  The radiation-axis proxy $\hat{\boldsymbol L}(t)$ traces a cone with half-angle $\theta_{\rm LJ}$ and phase $\Phi_{\rm p}$.
  Its direction is $(\theta_{\rm L},\varphi_{\rm L})$, with $\varphi_{\rm L}=\arg(\hat L_x+i\hat L_y)-\pi/2$ measured from $+\hat{\boldsymbol y}$; $\psi_{\rm L}$ completes the minimally rotating co-precessing frame.
  Angles and vector lengths are exaggerated.}
  \label{fig:frame_axis}
\end{figure}

The QA projection in \cref{eq:qa_projection} requires the radiation-axis direction $(\theta_{\rm L},\varphi_{\rm L})$ and a rotation $\psi_{\rm L}$ about that axis.
We reconstruct these angles from the waveform modes using the principal radiation axis \cite{OShaughnessy:2011pmr}, denoted $\hat{\boldsymbol L}$ as a waveform-based proxy for the orbital axis.
Acting on the strain multipoles with angular-momentum operators $L_a$ in the $|\ell m\rangle$ basis gives the real symmetric matrix
\begin{align}
  \Lambda_{ab}(t)
  =
  \sum_{\ell=2}^{\ell_{\rm ax}}
  \sum_{m,m'=-\ell}^{\ell}
  h_{\ell m}^{*}(t)\,
  \left\langle\ell m\left|
  \frac{L_aL_b+L_bL_a}{2}
  \right|\ell m'\right\rangle
  h_{\ell m'}(t),
\end{align}
with $a,b\in\{x,y,z\}$ and $\ell_{\rm ax}=4$.
Normalisation by the total mode power is unnecessary because it does not change the eigenvectors.
We identify $\hat{\boldsymbol L}(t)$ with the eigenvector of the largest eigenvalue, fix its sign by continuity from $+\hat{\boldsymbol z}$, and obtain
\begin{align}
  \theta_{\rm L}=\arccos\hat L_z,
  \qquad
  \varphi_{\rm L}=\arg\!\left(\hat L_x+i\hat L_y\right)-\frac{\pi}{2}.
  \label{eq:euler_from_axis}
\end{align}
We unwrap $\varphi_{\rm L}$ in time; the $-\pi/2$ shift sets $\varphi_{\rm L}=0$ for an axis in the $yz$ plane.
The minimal-rotation condition in \cref{eq:qa_projection}, with $\psi_{\rm L}(t_{\rm start})=0$, supplies the third angle.
\Cref{fig:frame_axis} illustrates these conventions.
For quadrupole-dominated quasicircular inspiral, the principal radiation axis approximately follows the instantaneous orbital angular momentum \cite{OShaughnessy:2011pmr}.

The reconstruction is well conditioned only when the largest eigenvalue is separated from the next.
Writing $\lambda_1\ge\lambda_2\ge\lambda_3$, we therefore monitor $(\lambda_1-\lambda_2)/\lambda_1$.
Near and after merger, $\lambda_1$ and $\lambda_2$ approach degeneracy, so both the axis and the co-precessing decomposition lose precision.
As a closure check, rotating the inertial modes into the QA frame and back recovers the input to near machine precision over the analysis window.
The resulting axis provides a waveform-based test of the spin-induced misalignment prescribed in the initial data; it should not be interpreted as an independent coordinate measure of the orbital plane near merger.

\begin{figure}
  \centering
  \includegraphics[width=\columnwidth]{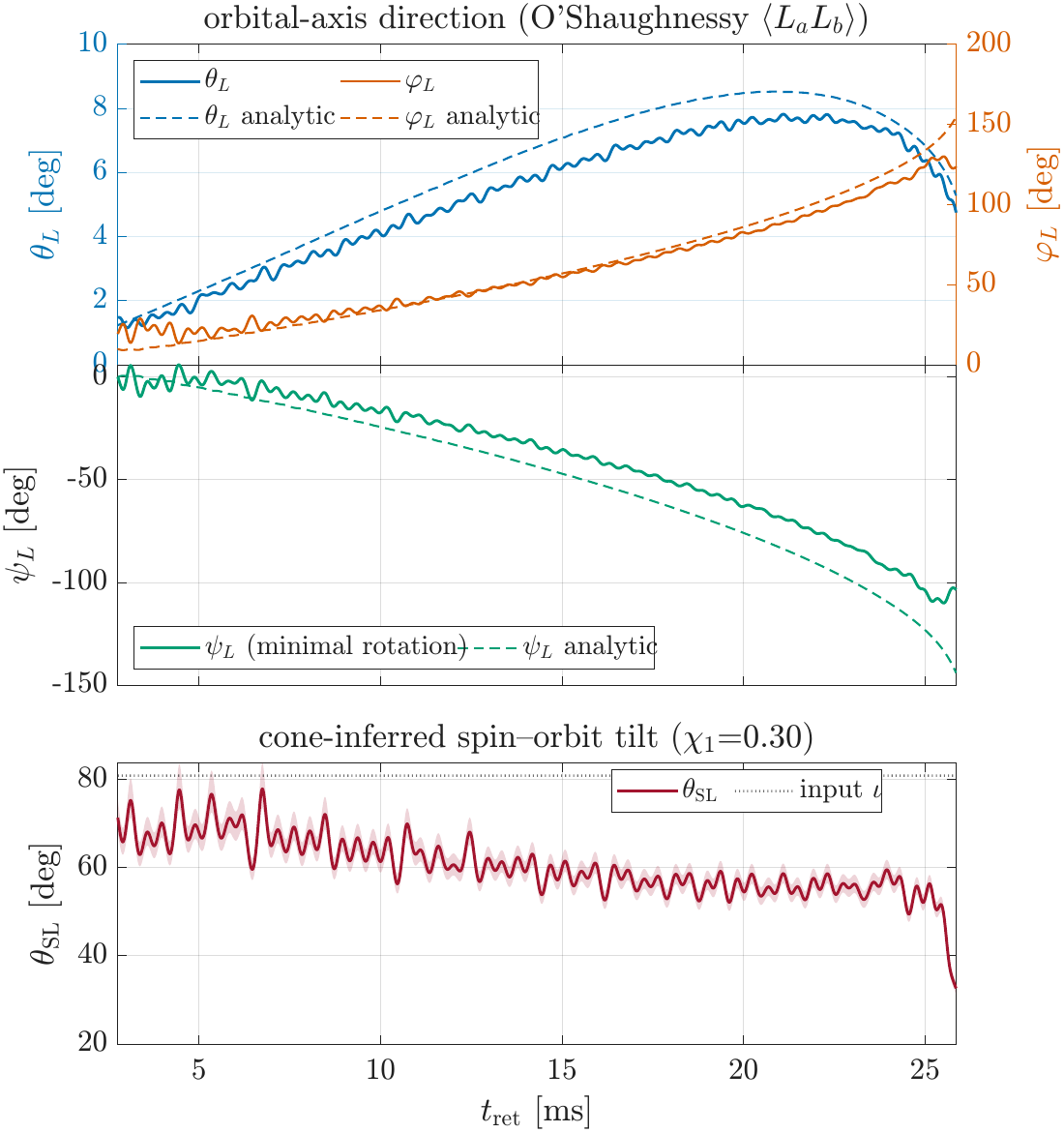}
  \caption{Radiation-axis precession reconstructed from the waveform of the same precessing BNS as in \cref{tab:benchmarks} and compared with the single-spin simple-precession model using a 3.5PN spin--orbit rate [dashed, \crefrange{eq:pn_L}{eq:Lhat_analytic}].
  \textbf{Top:} the polar tilt $\theta_{\rm L}$ and precession angle $\varphi_{\rm L}$ of $\hat{\boldsymbol L}(t)$.
  \textbf{Middle:} the minimal-rotation angle $\psi_{\rm L}$.
  The PN angles track the reconstruction through most of the inspiral and depart near merger ($t_{\rm ret}\gtrsim25$~ms), where the two leading eigenvalues of $\Lambda_{ab}$ approach degeneracy.
  \textbf{Bottom:} the inversion from \cref{eq:cone_inverse}, with a pointwise envelope obtained by propagating the combined statistical and subwindow-drift uncertainty in $\theta_{\rm LJ}$.
  A horizontal reference line marks the initial tilt.}
  \label{fig:orbital_axis}
\end{figure}

To test the reconstructed axis during inspiral, we compare it with the single-spin simple-precession model \cite{Apostolatos:1994mx} in the conventions of \cref{fig:frame_axis}.
The spinning star has $S_1=m_1^2\chi_1$ at tilt $\iota$ from the orbital axis.
The model assumes that $\hat{\boldsymbol L}$ precesses about the nearly conserved $\boldsymbol J=\boldsymbol L+\boldsymbol S_1$ while radiation reaction adiabatically decreases $L$.
It is therefore restricted to a quasicircular binary in the PN slow-motion regime, with well-separated orbital, precession, and radiation-reaction timescales \cite{Apostolatos:1994mx,Bohe:2012mr}.
The approximation becomes progressively less controlled during the strong-field late inspiral.
The nonspinning orbital angular momentum is given by
\begin{align}
  L(v)=\frac{\eta m^2}{v}\left[1+\left(\tfrac32+\tfrac{\eta}{6}\right)v^2
  +\left(\tfrac{27}{8}-\tfrac{19\eta}{8}+\tfrac{\eta^2}{24}\right)v^4\right],
  \label{eq:pn_L}
\end{align}
where $\eta=m_1m_2/m^2$ and $v=(m\Omega_{\rm orb})^{1/3}$ with $\Omega_{\rm orb}=\omega_{22}/2$ from the co-precessing $(2,2)$ frequency.
Because $\omega_{22}$ is measured from the numerical waveform rather than evolved by the PN model, the resulting angles test the precession dynamics conditional on the measured frequency; they do not independently predict the frequency evolution.
The vector relation $\boldsymbol J=\boldsymbol L+\boldsymbol S_1$ then fixes the cone half-angle,
\begin{align}
  \theta_{\rm LJ}(v)&=\arctan\frac{S_1\sin\iota}{L(v)+S_1\cos\iota}\,,
\end{align}
and total angular momentum,
\begin{align}
  J(v)&=\sqrt{L^2+2LS_1\cos\iota+S_1^2}\,,
\end{align}
respectively.
The spin-precession phase is $\Phi_{\rm p}(t)=\int_0^t \Omega_{\rm p}\,dt'$, where the precession frequency through 3.5PN order is~\cite{Bohe:2012mr}
\begin{align}
  \Omega_{\rm p}(v)=\frac{J(v)}{L(v)}\,\Omega_1(v)\,,
\end{align}
and the precession frequency of $\boldsymbol S_1$ about $\hat{\boldsymbol L}$ is
\begin{align}
  \Omega_1(v)=\frac{v^5}{m}\left(A_0+A_1v^2+A_2v^4\right)\,.
\end{align}
In above, the coefficients are
\begin{align}
  A_0&=\tfrac34+\tfrac{\eta}{2}-\tfrac34\delta,\qquad
  A_1=\tfrac{9}{16}+\tfrac54\eta-\tfrac{\eta^2}{24}
      +\delta\!\left(-\tfrac{9}{16}+\tfrac58\eta\right),\notag\\
  A_2&=\tfrac{27}{32}+\tfrac{3}{16}\eta-\tfrac{105}{32}\eta^2-\tfrac{\eta^3}{48}
      +\delta\!\left(-\tfrac{27}{32}+\tfrac{39}{8}\eta-\tfrac{5}{32}\eta^2\right)\,,
\end{align}
with $\delta=(m_1-m_2)/m$.
At leading order, this yields $\Omega_{\rm p}=(2+3m_2/2m_1)Jv^6/m^3$.
The predicted orbital axis therefore follows the cone as
\begin{align}
  \hat{\boldsymbol L}(t)=\cos\theta_{\rm LJ}\,\hat{\boldsymbol J}
  +\sin\theta_{\rm LJ}\left(\cos\Phi_{\rm p}\,\hat{\boldsymbol e}_1
  +\sin\Phi_{\rm p}\,\hat{\boldsymbol e}_2\right),
  \label{eq:Lhat_analytic}
\end{align}
where $\hat{\boldsymbol e}_1\propto\hat{\boldsymbol z} -(\hat{\boldsymbol J}\!\cdot\!\hat{\boldsymbol z})\,\hat{\boldsymbol J}$ and $\hat{\boldsymbol e}_2=\hat{\boldsymbol J}\times\hat{\boldsymbol e}_1$ are fixed so that $\hat{\boldsymbol L}\parallel\hat{\boldsymbol z}$ at $t=0$.
Substituting $\hat{\boldsymbol L}(t)$ into \cref{eq:euler_from_axis} gives $\theta_{\rm L}(t)$ and $\varphi_{\rm L}(t)$, and integrating the minimal-rotation condition in \cref{eq:qa_projection} determines $\psi_{\rm L}(t)$.
These three predicted angles are shown by the dashed curves in \cref{fig:orbital_axis}.

It is however not easy to test the initial spin--orbit tilt $\iota$ from the waveform.
Within the simple-precession approximation, the angle between $\boldsymbol S_1$ and $\boldsymbol L$ ($\theta_{\rm SL}$) remains constant and equals the initial tilt $\iota=:\theta_{\rm SL}(t=0)$.
Rearranging the cone relation $\sin(\theta_{\rm SL}-\theta_{\rm LJ})=(L/S_1)\sin\theta_{\rm LJ}$ gives
\begin{align}
  \theta_{\rm SL} = \theta_{\rm LJ}+\arcsin[(L/S_1)\sin\theta_{\rm LJ}].
  \label{eq:cone_inverse}
\end{align}
This inversion is ill conditioned for a nearly in-plane spin.
The forward map $\tan\theta_{\rm LJ}=S_1\sin\theta_{\rm SL}/(L+S_1\cos\theta_{\rm SL})$ reaches $\theta_{\rm LJ}^{\max}=\arcsin(S_1/L)$ at $\theta_{\rm SL}^{*}=\arccos(-S_1/L)$, where $\mathrm{d}\theta_{\rm LJ}/\mathrm{d}\theta_{\rm SL}$ vanishes.
For the present configuration $L/S_1\approx13.8$, giving $\theta_{\rm LJ}^{\max}\approx4.2^\circ$ at $\theta_{\rm SL}^{*}\approx94^\circ$.
Here $\mathrm{d}\theta_{\rm SL}/\mathrm{d}\theta_{\rm LJ}\approx26$, so small errors in the cone angle are strongly amplified.
The ill-conditioned inversion for this nearly in-plane spin gives about $60^\circ$, below the input $81^\circ$.
A complementary estimate follows from the initial single-spin balance
\begin{align}
  \cos\theta_{\rm SL}=\frac{J_{z,0}-L_0}{S_1}\,,
  \label{eq:tilt_balance}
\end{align}
where $J_{z,0}$ is the initial ADM angular momentum along the orbital axis and $L_0$ is the PN orbital contribution at the same reference time.
The angular-momentum-balance estimate from \cref{eq:tilt_balance} gives about $82^\circ$, which is more consistent with the tilt of initial data.

\section{Conclusion}
\label{sec:conclusion}
\celephais provides spectrally resolved compact-binary initial data without restricting the spins to align or antialign with the orbital axis.
Its central numerical result relies on an equation-to-Jacobian machinery that exploits sparsity, and the fact that an intermittently refreshed direct preconditioner is enough to guarantee convergence.
In particular, the operator-tree filter removes structural zeros, batched forward-mode automatic differentiation evaluates compatible columns together, and \texttt{MUMPS} factors the assembled matrix for use in a Jacobian-free Newton--Krylov solve.
Together, the benchmarks show that the precessing BNS can be solved on a laptop in a matter of minutes (\cref{tab:benchmarks}).

In addition, we implement an AMR algorithm aiming to ease the resource demand to solve for high-mass ratio BHNS.
For the mass-ratio-$20$ BHNS, adaptive refinement reproduces the highest uniform-grid constraint accuracy with about three times fewer unknowns (\cref{fig:amr_cost_q20}), while $p$ refinement reaches the uniform $N=15$ accuracy for the mass-ratio-$10$ BHNS with about $1.6$ times fewer unknowns (\cref{fig:spectral_tail}).
These gains arise because computational effort is concentrated on the couplings, domains, and spectral directions that limit convergence.

We first validate \celephais by importing solutions generated with \texttt{FUKA} and confirming that each remains a solution of the corresponding system in \celephais.
We then selectively build some BNS and BHNS sequences to examine the consistency with post-Newtonian theory, where the comparatively small spin-dependent effects can also be well captured (\cref{fig:pn}).
This agreement motivates the use of post-Newtonian predictions for the orbital frequency and infall velocity to construct low-eccentricity configurations, following the approach of \cite{Papenfort:2021hod} but including spin contributions here.
This scheme allows for constructing quasi-circular precessing BNS by running only one round of evolution-based correction (\cref{fig:ecc}).
The precessing BNS is then evolved stably through inspiral and merger to verify that the initial data produce a reliable gravitational waveform.
In particular, the radiation axis reconstructed from the waveform closely follows the 3.5PN simple-precession model throughout most of the inspiral (\cref{fig:orbital_axis}).
Taken together, these tests link the spin geometry prescribed in the initial data to the expected orbital dynamics and waveform evolution.

The evolution presented here demonstrates usability and provides the multipoles needed for radiation-axis reconstruction, but it does not establish waveform-phase convergence. 
Such a study requires multiple resolutions, careful estimator selection~\cite{Kuan:2024jnw,Kuan:2025bzu,Kiuchi:2025ksk,2014JCoPh.262..104E,2014PhFl...26c5101O}, and the self-similar scaling test of phase errors \cite{Dietrich:2019kaq,Hotokezaka:2015xka}.
Precession adds further difficulty because meaningful phase comparisons require a time-dependent co-precessing frame. 
We therefore defer the detailed waveform analysis to future work. 
In addition, the black-hole is treated with an excision horizon with. 
A puncture formulation is planned to allow for initial data matching with the moving-puncture gauge commonly used in NR simulations.
Extensions to Damour--Esposito-Far\`ese scalar--tensor (as an upgrade of \cite{Kuan:2023hrh}) and scalar Gauss--Bonnet gravity are also planned.

\section*{Acknowledgements}
The author acknowledges support from the Simons Foundation through Award No.~896696, Simons Foundation International through Award No.~SFI-MPS-BH-00012593-01, and the NSF through Grant No.~PHY-25-12423.
The author thanks Kenta Kiuchi for encouraging the development of an initial-data solver.
Yong Gao, Mingzhe Han, and Alan Tsz-Lok Lam assisted with the numerical evolutions.
Computations were performed in part on the BinAC2 cluster, supported by the High Performance and Cloud Computing Group at the Zentrum f{\"u}r Datenverarbeitung of the University of T{\"u}bingen, and, during the early stages of the project, on the Sakura cluster at the Max Planck Computing and Data Facility.
Claude was used extensively to assist the analysis of numerical bottlenecks and optimisation strategies.

\section*{Code availability}
\celephais will be released under the GNU General Public License upon article acceptance.
The release will include the parameters needed to reproduce the configurations and figures, together with the \textsc{Matlab} analysis scripts.

\bibliography{inspire,NOTinspire}

@misc{LORENE,
    title        = {{LORENE website}: {L}angage {O}bjet pour la {RE}lativit{\'e}
                    {N}um{\'e}riqu{\'e}},
    howpublished = {\url{https://gitlab.in2p3.fr/lorene/Lorene}}
}

@article{MUMPS,
  author = {Amestoy, Patrick R. and Duff, Iain S. and L'Excellent, Jean-Yves and Koster, Jacko},
  title = {A Fully Asynchronous Multifrontal Solver Using Distributed Dynamic Scheduling},
  journal = {SIAM Journal on Matrix Analysis and Applications},
  volume = {23},
  number = {1},
  pages = {15-41},
  year = {2001},
  doi = {10.1137/S0895479899358194},
  eprint = {https://doi.org/10.1137/S0895479899358194}
}

@book{Griewank:2008ed,
  author    = {Griewank, Andreas and Walther, Andrea},
  title     = {Evaluating Derivatives: Principles and Techniques of Algorithmic Differentiation},
  edition   = {2},
  publisher = {Society for Industrial and Applied Mathematics},
  address   = {Philadelphia},
  year      = {2008},
  doi       = {10.1137/1.9780898717761},
  isbn      = {978-0-89871-659-7}
}

@article{EisenstatWalker1996,
  author = {Eisenstat, Stanley C. and Walker, Homer F.},
  title = {Choosing the Forcing Terms in an Inexact Newton Method},
  journal = {SIAM Journal on Scientific Computing},
  volume = {17},
  number = {1},
  pages = {16-32},
  year = {1996},
  doi = {10.1137/0917003},
  URL = {https://doi.org/10.1137/0917003},
  eprint = {https://doi.org/10.1137/0917003}
}

@article{DemboEisenstatSteihaug1982,
  author = {Dembo, Ron S. and Eisenstat, Stanley C. and Steihaug, Trond},
  title = {Inexact Newton Methods},
  journal = {SIAM Journal on Numerical Analysis},
  volume = {19},
  number = {2},
  pages = {400-408},
  year = {1982},
  doi = {10.1137/0719025},
  URL = {https://doi.org/10.1137/0719025},
  eprint = {https://doi.org/10.1137/0719025}
}

@inproceedings{PerssonPeraire:2006,
  author    = {Persson, Per-Olof and Peraire, Jaime},
  title     = {Sub-Cell Shock Capturing for Discontinuous Galerkin Methods},
  booktitle = {44th AIAA Aerospace Sciences Meeting and Exhibit},
  pages     = {112},
  year      = {2006},
  publisher = {American Institute of Aeronautics and Astronautics},
  doi       = {10.2514/6.2006-112}
}

@article{Doerfler:1996,
  author = {D{\"o}rfler, Willy},
  title = {A Convergent Adaptive Algorithm for Poisson’s Equation},
  journal = {SIAM Journal on Numerical Analysis},
  volume = {33},
  number = {3},
  pages = {1106-1124},
  year = {1996},
  doi = {10.1137/0733054},
  URL = {https://doi.org/10.1137/0733054},
  eprint = {https://doi.org/10.1137/0733054},
}

@ARTICLE{2014PhFl...26c5101O,
       author = {{Oliver}, Todd A. and {Malaya}, Nicholas and {Ulerich}, Rhys and {Moser}, Robert D.},
        title = "{Estimating uncertainties in statistics computed from direct numerical simulation}",
      journal = {Physics of Fluids},
         year = 2014,
        month = mar,
       volume = {26},
       number = {3},
          eid = {035101},
        pages = {035101},
          doi = {10.1063/1.4866813},
archivePrefix = {arXiv},
       eprint = {1311.0828},
 primaryClass = {physics.flu-dyn},
       adsurl = {https://ui.adsabs.harvard.edu/abs/2014PhFl...26c5101O}
}

@ARTICLE{2014JCoPh.262..104E,
       author = {{E{\c{c}}a}, L. and {Hoekstra}, M.},
        title = "{A procedure for the estimation of the numerical uncertainty of CFD calculations based on grid refinement studies}",
      journal = {Journal of Computational Physics},
         year = 2014,
        month = apr,
       volume = {262},
        pages = {104-130},
          doi = {10.1016/j.jcp.2014.01.006},
       adsurl = {https://ui.adsabs.harvard.edu/abs/2014JCoPh.262..104E}
}

@ARTICLE{1998SJSC...20..359K,
       author = {{Karypis}, George and {Kumar}, Vipin},
        title = "{A Fast and High Quality Multilevel Scheme for Partitioning Irregular Graphs}",
      journal = {SIAM Journal on Scientific Computing},
         year = 1998,
        month = jan,
       volume = {20},
       number = {1},
        pages = {359-392},
          doi = {10.1137/S1064827595287997},
       adsurl = {https://ui.adsabs.harvard.edu/abs/1998SJSC...20..359K}
}

@article{KAGRA:2021vkt,
    author = "Abbott, R. and others",
    collaboration = "KAGRA, VIRGO, LIGO Scientific",
    title = "{GWTC-3: Compact Binary Coalescences Observed by LIGO and Virgo during the Second Part of the Third Observing Run}",
    eprint = "2111.03606",
    archivePrefix = "arXiv",
    primaryClass = "gr-qc",
    reportNumber = "LIGO-P2000318",
    doi = "10.1103/PhysRevX.13.041039",
    journal = "Phys. Rev. X",
    volume = "13",
    number = "4",
    pages = "041039",
    year = "2023"
}

@article{Hotokezaka:2015xka,
    author = "Hotokezaka, Kenta and Kyutoku, Koutarou and Okawa, Hirotada and Shibata, Masaru",
    title = "{Exploring tidal effects of coalescing binary neutron stars in numerical relativity. II. Long-term simulations}",
    eprint = "1502.03457",
    archivePrefix = "arXiv",
    primaryClass = "gr-qc",
    doi = "10.1103/PhysRevD.91.064060",
    journal = "Phys. Rev. D",
    volume = "91",
    number = "6",
    pages = "064060",
    year = "2015"
}

@article{Kawaguchi:2017wrt,
    author = "Kawaguchi, Kyohei and Kyutoku, Koutarou and Nakano, Hiroyuki and Shibata, Masaru",
    title = "{Extracting the orbital axis from gravitational waves of precessing binary systems}",
    eprint = "1705.07459",
    archivePrefix = "arXiv",
    primaryClass = "gr-qc",
    doi = "10.1103/PhysRevD.97.024017",
    journal = "Phys. Rev. D",
    volume = "97",
    number = "2",
    pages = "024017",
    year = "2018"
}

@inproceedings{York:1978gql,
    author = "York, Jr., James W.",
    title = "{Kinematics and Dynamics of General Relativity}",
    booktitle = "{Workshop on Sources of Gravitational Radiation}",
    pages = "83--126",
    year = "1978"
}

@article{OMurchadha:1974pq,
    author = "O Murchadha, Niall and York, James W.",
    title = "{Gravitational energy}",
    doi = "10.1103/PhysRevD.10.2345",
    journal = "Phys. Rev. D",
    volume = "10",
    pages = "2345--2357",
    year = "1974"
}

@article{Jaramillo:2004uc,
    author = "Jaramillo, Jose Luis and Gourgoulhon, Eric and Mena Marugan, Guillermo A.",
    title = "{Inner boundary conditions for black hole initial data derived from isolated horizons}",
    eprint = "gr-qc/0407063",
    archivePrefix = "arXiv",
    doi = "10.1103/PhysRevD.70.124036",
    journal = "Phys. Rev. D",
    volume = "70",
    pages = "124036",
    year = "2004"
}

@article{Kiendrebeogo:2023hzf,
    author = "Kiendrebeogo, R. Weizmann and others",
    title = "{Updated Observing Scenarios and Multimessenger Implications for the International Gravitational-wave Networks O4 and O5}",
    eprint = "2306.09234",
    archivePrefix = "arXiv",
    primaryClass = "astro-ph.HE",
    doi = "10.3847/1538-4357/acfcb1",
    journal = "Astrophys. J.",
    volume = "958",
    number = "2",
    pages = "158",
    year = "2023"
}

@article{Bonazzola:1998ge,
    author = "Bonazzola, S. and Gourgoulhon, E. and Marck, J. A.",
    title = "{Spectral methods in general relativistic astrophysics}",
    eprint = "gr-qc/9811089",
    archivePrefix = "arXiv",
    doi = "10.1016/S0377-0427(99)00167-3",
    journal = "J. Comput. Appl. Math.",
    volume = "109",
    pages = "433",
    year = "1999"
}

@article{Grandclement:2007sb,
    author = "Grandclement, Philippe and Novak, Jerome",
    title = "{Spectral methods for numerical relativity}",
    eprint = "0706.2286",
    archivePrefix = "arXiv",
    primaryClass = "gr-qc",
    doi = "10.12942/lrr-2009-1",
    journal = "Living Rev. Rel.",
    volume = "12",
    pages = "1",
    year = "2009"
}

@article{Gourgoulhon:2000nn,
    author = "Gourgoulhon, Eric and Grandclement, Philippe and Taniguchi, Keisuke and Marck, Jean-Alain and Bonazzola, Silvano",
    title = "{Quasiequilibrium sequences of synchronized and irrotational binary neutron stars in general relativity: 1. Method and tests}",
    eprint = "gr-qc/0007028",
    archivePrefix = "arXiv",
    doi = "10.1103/PhysRevD.63.064029",
    journal = "Phys. Rev. D",
    volume = "63",
    pages = "064029",
    year = "2001"
}

@article{Grandclement:2001ed,
    author = "Grandclement, Philippe and Gourgoulhon, Eric and Bonazzola, Silvano",
    title = "{Binary black holes in circular orbits. 2. Numerical methods and first results}",
    eprint = "gr-qc/0106016",
    archivePrefix = "arXiv",
    doi = "10.1103/PhysRevD.65.044021",
    journal = "Phys. Rev. D",
    volume = "65",
    pages = "044021",
    year = "2002"
}

@article{Grandclement:2009ju,
    author = "Grandclement, Philippe",
    title = "{Kadath: A Spectral solver for theoretical physics}",
    eprint = "0909.1228",
    archivePrefix = "arXiv",
    primaryClass = "gr-qc",
    doi = "10.1016/j.jcp.2010.01.005",
    journal = "J. Comput. Phys.",
    volume = "229",
    pages = "3334--3357",
    year = "2010"
}

@article{Papenfort:2021hod,
    author = "Papenfort, L. Jens and Tootle, Samuel D. and Grandcl{\'e}ment, Philippe and Most, Elias R. and Rezzolla, Luciano",
    title = "{New public code for initial data of unequal-mass, spinning compact-object binaries}",
    eprint = "2103.09911",
    archivePrefix = "arXiv",
    primaryClass = "gr-qc",
    doi = "10.1103/PhysRevD.104.024057",
    journal = "Phys. Rev. D",
    volume = "104",
    number = "2",
    pages = "024057",
    year = "2021"
}

@article{Dietrich:2015pxa,
    author = {Dietrich, Tim and Moldenhauer, Niclas and Johnson-McDaniel, Nathan K. and Bernuzzi, Sebastiano and Markakis, Charalampos M. and Br{\"u}gmann, Bernd and Tichy, Wolfgang},
    title = "{Binary Neutron Stars with Generic Spin, Eccentricity, Mass ratio, and Compactness - Quasi-equilibrium Sequences and First Evolutions}",
    eprint = "1507.07100",
    archivePrefix = "arXiv",
    primaryClass = "gr-qc",
    reportNumber = "ICTS-2015-5",
    doi = "10.1103/PhysRevD.92.124007",
    journal = "Phys. Rev. D",
    volume = "92",
    number = "12",
    pages = "124007",
    year = "2015"
}

@article{Tichy:2019ouu,
    author = {Tichy, Wolfgang and Rashti, Alireza and Dietrich, Tim and Dudi, Reetika and Br{\"u}gmann, Bernd},
    title = "{Constructing binary neutron star initial data with high spins, high compactnesses, and high mass ratios}",
    eprint = "1910.09690",
    archivePrefix = "arXiv",
    primaryClass = "gr-qc",
    doi = "10.1103/PhysRevD.100.124046",
    journal = "Phys. Rev. D",
    volume = "100",
    number = "12",
    pages = "124046",
    year = "2019"
}

@article{Rashti:2021ihv,
    author = {Rashti, Alireza and Fabbri, Francesco Maria and Br{\"u}gmann, Bernd and Chaurasia, Swami Vivekanandji and Dietrich, Tim and Ujevic, Maximiliano and Tichy, Wolfgang},
    title = "{New pseudospectral code for the construction of initial data}",
    eprint = "2109.14511",
    archivePrefix = "arXiv",
    primaryClass = "gr-qc",
    doi = "10.1103/PhysRevD.105.104027",
    journal = "Phys. Rev. D",
    volume = "105",
    number = "10",
    pages = "104027",
    year = "2022"
}

@article{Pfeiffer:2002wt,
    author = "Pfeiffer, Harald P. and Kidder, Lawrence E. and Scheel, Mark A. and Teukolsky, Saul A.",
    title = "{A Multidomain spectral method for solving elliptic equations}",
    eprint = "gr-qc/0202096",
    archivePrefix = "arXiv",
    doi = "10.1016/S0010-4655(02)00847-0",
    journal = "Comput. Phys. Commun.",
    volume = "152",
    pages = "253--273",
    year = "2003"
}

@article{Foucart:2008qt,
    author = "Foucart, Francois and Kidder, Lawrence E. and Pfeiffer, Harald P. and Teukolsky, Saul A.",
    title = "{Initial data for black hole-neutron star binaries: A Flexible, high-accuracy spectral method}",
    eprint = "0804.3787",
    archivePrefix = "arXiv",
    primaryClass = "gr-qc",
    doi = "10.1103/PhysRevD.77.124051",
    journal = "Phys. Rev. D",
    volume = "77",
    pages = "124051",
    year = "2008"
}

@article{Tacik:2015tja,
    author = "Tacik, Nick and others",
    title = "{Binary Neutron Stars with Arbitrary Spins in Numerical Relativity}",
    eprint = "1508.06986",
    archivePrefix = "arXiv",
    primaryClass = "gr-qc",
    doi = "10.1103/PhysRevD.92.124012",
    journal = "Phys. Rev. D",
    volume = "92",
    number = "12",
    pages = "124012",
    year = "2015",
    note = "[Erratum: Phys.Rev.D 94, 049903 (2016)]"
}

@article{Assumpcao:2021fhq,
    author = "Assumpcao, Thiago and Werneck, Leonardo R. and Jacques, Terrence Pierre and Etienne, Zachariah B.",
    title = "{Fast hyperbolic relaxation elliptic solver for numerical relativity: Conformally flat, binary puncture initial data}",
    eprint = "2111.02424",
    archivePrefix = "arXiv",
    primaryClass = "gr-qc",
    doi = "10.1103/PhysRevD.105.104037",
    journal = "Phys. Rev. D",
    volume = "105",
    number = "10",
    pages = "104037",
    year = "2022"
}

@article{Vu:2021coj,
    author = "Vu, Nils L. and others",
    title = "{A scalable elliptic solver with task-based parallelism for the SpECTRE numerical relativity code}",
    eprint = "2111.06767",
    archivePrefix = "arXiv",
    primaryClass = "gr-qc",
    doi = "10.1103/PhysRevD.105.084027",
    journal = "Phys. Rev. D",
    volume = "105",
    number = "8",
    pages = "084027",
    year = "2022"
}

@article{Vincent:2019qpd,
    author = "Vincent, Trevor and Pfeiffer, Harald P. and Fischer, Nils L.",
    title = "{hp-adaptive discontinuous Galerkin solver for elliptic equations in numerical relativity}",
    eprint = "1907.01572",
    archivePrefix = "arXiv",
    primaryClass = "physics.comp-ph",
    doi = "10.1103/PhysRevD.100.084052",
    journal = "Phys. Rev. D",
    volume = "100",
    number = "8",
    pages = "084052",
    year = "2019"
}

@misc{Gourgoulhon:2007ue,
    author = "Gourgoulhon, Eric",
    title = "{3+1 formalism and bases of numerical relativity}",
    eprint = "gr-qc/0703035",
    archivePrefix = "arXiv",
    month = "3",
    year = "2007"
}

@article{York:1998hy,
    author = "York, Jr., James W.",
    title = "{Conformal 'thin sandwich' data for the initial-value problem}",
    eprint = "gr-qc/9810051",
    archivePrefix = "arXiv",
    reportNumber = "IFP-UNC-527",
    doi = "10.1103/PhysRevLett.82.1350",
    journal = "Phys. Rev. Lett.",
    volume = "82",
    pages = "1350--1353",
    year = "1999"
}

@article{Pfeiffer:2002iy,
    author = "Pfeiffer, Harald P. and York, Jr., James W.",
    title = "{Extrinsic curvature and the Einstein constraints}",
    eprint = "gr-qc/0207095",
    archivePrefix = "arXiv",
    doi = "10.1103/PhysRevD.67.044022",
    journal = "Phys. Rev. D",
    volume = "67",
    pages = "044022",
    year = "2003"
}

@article{Cook:2001wi,
    author = "Cook, Gregory B.",
    title = "{Corotating and irrotational binary black holes in quasicircular orbits}",
    eprint = "gr-qc/0108076",
    archivePrefix = "arXiv",
    doi = "10.1103/PhysRevD.65.084003",
    journal = "Phys. Rev. D",
    volume = "65",
    pages = "084003",
    year = "2002"
}

@article{Bonazzola:2003dm,
    author = "Bonazzola, Silvano and Gourgoulhon, Eric and Grandclement, Philippe and Novak, Jerome",
    title = "{A Constrained scheme for Einstein equations based on Dirac gauge and spherical coordinates}",
    eprint = "gr-qc/0307082",
    archivePrefix = "arXiv",
    doi = "10.1103/PhysRevD.70.104007",
    journal = "Phys. Rev. D",
    volume = "70",
    pages = "104007",
    year = "2004"
}

@article{Isenberg:2007zg,
    author = "Isenberg, James A.",
    title = "{Waveless approximation theories of gravity}",
    eprint = "gr-qc/0702113",
    archivePrefix = "arXiv",
    doi = "10.1142/S0218271808011997",
    journal = "Int. J. Mod. Phys. D",
    volume = "17",
    pages = "265--273",
    year = "2008"
}

@article{Wilson:1995uh,
    author = "Wilson, J. R. and Mathews, G. J.",
    title = "{Instabilities in Close Neutron Star Binaries}",
    doi = "10.1103/PhysRevLett.75.4161",
    journal = "Phys. Rev. Lett.",
    volume = "75",
    pages = "4161--4164",
    year = "1995"
}

@article{Wilson:1996ty,
    author = "Wilson, J. R. and Mathews, G. J. and Marronetti, P.",
    title = "{Relativistic numerical model for close neutron star binaries}",
    eprint = "gr-qc/9601017",
    archivePrefix = "arXiv",
    doi = "10.1103/PhysRevD.54.1317",
    journal = "Phys. Rev. D",
    volume = "54",
    pages = "1317--1331",
    year = "1996"
}

@article{Shibata:2004qz,
    author = "Shibata, Masaru and Uryu, Koji and Friedman, John L.",
    title = "{Deriving formulations for numerical computation of binary neutron stars in quasicircular orbits}",
    eprint = "gr-qc/0407036",
    archivePrefix = "arXiv",
    doi = "10.1103/PhysRevD.70.044044",
    journal = "Phys. Rev. D",
    volume = "70",
    pages = "044044",
    year = "2004",
    note = "[Erratum: Phys.Rev.D 70, 129901 (2004)]"
}

@article{Uryu:2005vv,
    author = "Uryu, Koji and Limousin, Francois and Friedman, John L. and Gourgoulhon, Eric and Shibata, Masaru",
    title = "{Binary neutron stars in a waveless approximation}",
    eprint = "gr-qc/0511136",
    archivePrefix = "arXiv",
    doi = "10.1103/PhysRevLett.97.171101",
    journal = "Phys. Rev. Lett.",
    volume = "97",
    pages = "171101",
    year = "2006"
}

@article{Bonazzola:1997gc,
    author = "Bonazzola, Silvano and Gourgoulhon, Eric and Marck, Jean-Alain",
    title = "{A Relativistic formalism to compute quasiequilibrium configurations of nonsynchronized neutron star binaries}",
    eprint = "gr-qc/9710031",
    archivePrefix = "arXiv",
    doi = "10.1103/PhysRevD.56.7740",
    journal = "Phys. Rev. D",
    volume = "56",
    pages = "7740--7749",
    year = "1997"
}

@article{Tichy:2003zg,
    author = "Tichy, Wolfgang and Bruegmann, Bernd and Laguna, Pablo",
    title = "{Gauge conditions for binary black hole puncture data based on an approximate helical Killing vector}",
    eprint = "gr-qc/0306020",
    archivePrefix = "arXiv",
    doi = "10.1103/PhysRevD.68.064008",
    journal = "Phys. Rev. D",
    volume = "68",
    pages = "064008",
    year = "2003"
}

@article{Ossokine:2015yla,
    author = "Ossokine, Serguei and Foucart, Francois and Pfeiffer, Harald P. and Boyle, Michael and Szil{\'a}gyi, B{\'e}la",
    title = "{Improvements to the construction of binary black hole initial data}",
    eprint = "1506.01689",
    archivePrefix = "arXiv",
    primaryClass = "gr-qc",
    doi = "10.1088/0264-9381/32/24/245010",
    journal = "Class. Quant. Grav.",
    volume = "32",
    pages = "245010",
    year = "2015"
}

@article{Moldenhauer:2014yaa,
    author = {Moldenhauer, Niclas and Markakis, Charalampos M. and Johnson-McDaniel, Nathan K. and Tichy, Wolfgang and Br{\"u}gmann, Bernd},
    title = "{Initial data for binary neutron stars with adjustable eccentricity}",
    eprint = "1408.4136",
    archivePrefix = "arXiv",
    primaryClass = "gr-qc",
    doi = "10.1103/PhysRevD.90.084043",
    journal = "Phys. Rev. D",
    volume = "90",
    number = "8",
    pages = "084043",
    year = "2014"
}

@article{Buonanno:2010yk,
    author = "Buonanno, Alessandra and Kidder, Lawrence E. and Mroue, Abdul H. and Pfeiffer, Harald P. and Taracchini, Andrea",
    title = "{Reducing orbital eccentricity of precessing black-hole binaries}",
    eprint = "1012.1549",
    archivePrefix = "arXiv",
    primaryClass = "gr-qc",
    doi = "10.1103/PhysRevD.83.104034",
    journal = "Phys. Rev. D",
    volume = "83",
    pages = "104034",
    year = "2011"
}

@article{Tichy:2009yr,
    author = "Tichy, Wolfgang",
    title = "{A New numerical method to construct binary neutron star initial data}",
    eprint = "0908.0620",
    archivePrefix = "arXiv",
    primaryClass = "gr-qc",
    doi = "10.1088/0264-9381/26/17/175018",
    journal = "Class. Quant. Grav.",
    volume = "26",
    pages = "175018",
    year = "2009"
}

@article{Tichy:2011gw,
    author = "Tichy, Wolfgang",
    title = "{Initial data for binary neutron stars with arbitrary spins}",
    eprint = "1107.1440",
    archivePrefix = "arXiv",
    primaryClass = "gr-qc",
    doi = "10.1103/PhysRevD.84.024041",
    journal = "Phys. Rev. D",
    volume = "84",
    pages = "024041",
    year = "2011"
}

@article{Tichy:2012rp,
    author = "Tichy, Wolfgang",
    title = "{Constructing quasi-equilibrium initial data for binary neutron stars with arbitrary spins}",
    eprint = "1209.5336",
    archivePrefix = "arXiv",
    primaryClass = "gr-qc",
    doi = "10.1103/PhysRevD.86.064024",
    journal = "Phys. Rev. D",
    volume = "86",
    pages = "064024",
    year = "2012"
}

@article{Shibata:1998um,
    author = "Shibata, Masaru",
    title = "{A Relativistic formalism for computation of irrotational binary stars in quasiequilibrium states}",
    eprint = "gr-qc/9803085",
    archivePrefix = "arXiv",
    reportNumber = "OU-TAP-73",
    doi = "10.1103/PhysRevD.58.024012",
    journal = "Phys. Rev. D",
    volume = "58",
    pages = "024012",
    year = "1998"
}

@misc{Gourgoulhon:1998dr,
    author = "Gourgoulhon, Eric",
    title = "{Relations between three formalisms for irrotational binary neutron stars in general relativity}",
    eprint = "gr-qc/9804054",
    archivePrefix = "arXiv",
    month = "4",
    year = "1998"
}

@article{Bonazzola:1998yq,
    author = "Bonazzola, Silvano and Gourgoulhon, Eric and Marck, Jean-Alain",
    title = "{Numerical models of irrotational binary neutron stars in general relativity}",
    eprint = "gr-qc/9810072",
    archivePrefix = "arXiv",
    doi = "10.1103/PhysRevLett.82.892",
    journal = "Phys. Rev. Lett.",
    volume = "82",
    pages = "892--895",
    year = "1999"
}

@article{Foucart:2010eq,
    author = "Foucart, Francois and Duez, Matthew D. and Kidder, Lawrence E. and Teukolsky, Saul A.",
    title = "{Black hole-neutron star mergers: effects of the orientation of the black hole spin}",
    eprint = "1007.4203",
    archivePrefix = "arXiv",
    primaryClass = "astro-ph.HE",
    doi = "10.1103/PhysRevD.83.024005",
    journal = "Phys. Rev. D",
    volume = "83",
    pages = "024005",
    year = "2011"
}

@article{Foucart:2012vn,
    author = "Foucart, Francois and Deaton, M. Brett and Duez, Matthew D. and Kidder, Lawrence E. and MacDonald, Ilana and Ott, Christian D. and Pfeiffer, Harald P. and Scheel, Mark A. and Szilagyi, Bela and Teukolsky, Saul A.",
    title = "{Black hole-neutron star mergers at realistic mass ratios: Equation of state and spin orientation effects}",
    eprint = "1212.4810",
    archivePrefix = "arXiv",
    primaryClass = "gr-qc",
    doi = "10.1103/PhysRevD.87.084006",
    journal = "Phys. Rev. D",
    volume = "87",
    pages = "084006",
    year = "2013"
}

@article{Tsatsin:2013jca,
    author = "Tsatsin, Petr and Marronetti, Pedro",
    title = "{Initial data for neutron star binaries with arbitrary spins}",
    eprint = "1303.6692",
    archivePrefix = "arXiv",
    primaryClass = "gr-qc",
    doi = "10.1103/PhysRevD.88.064060",
    journal = "Phys. Rev. D",
    volume = "88",
    number = "6",
    pages = "064060",
    year = "2013"
}

@article{Kawaguchi:2015bwa,
    author = "Kawaguchi, Kyohei and Kyutoku, Koutarou and Nakano, Hiroyuki and Okawa, Hirotada and Shibata, Masaru and Taniguchi, Keisuke",
    title = "{Black hole-neutron star binary merger: Dependence on black hole spin orientation and equation of state}",
    eprint = "1506.05473",
    archivePrefix = "arXiv",
    primaryClass = "astro-ph.HE",
    doi = "10.1103/PhysRevD.92.024014",
    journal = "Phys. Rev. D",
    volume = "92",
    number = "2",
    pages = "024014",
    year = "2015"
}

@article{Dietrich:2017xqb,
    author = {Dietrich, Tim and Bernuzzi, Sebastiano and Br{\"u}gmann, Bernd and Ujevic, Maximiliano and Tichy, Wolfgang},
    title = "{Numerical Relativity Simulations of Precessing Binary Neutron Star Mergers}",
    eprint = "1712.02992",
    archivePrefix = "arXiv",
    primaryClass = "gr-qc",
    doi = "10.1103/PhysRevD.97.064002",
    journal = "Phys. Rev. D",
    volume = "97",
    number = "6",
    pages = "064002",
    year = "2018"
}

@article{Kyutoku:2020xka,
    author = "Kyutoku, Koutarou and Fujibayashi, Sho and Hayashi, Kota and Kawaguchi, Kyohei and Kiuchi, Kenta and Shibata, Masaru and Tanaka, Masaomi",
    title = "{On the Possibility of GW190425 Being a Black Hole{\textendash}Neutron Star Binary Merger}",
    eprint = "2001.04474",
    archivePrefix = "arXiv",
    primaryClass = "astro-ph.HE",
    doi = "10.3847/2041-8213/ab6e70",
    journal = "Astrophys. J. Lett.",
    volume = "890",
    number = "1",
    pages = "L4",
    year = "2020"
}

@article{Bonazzola:1993zz,
    author = "Bonazzola, S. and Gourgoulhon, E. and Salgado, M. and Marck, J. A.",
    title = "{Axisymmetric rotating relativistic bodies: A new numerical approach for 'exact' solutions}",
    journal = "Astron. Astrophys.",
    volume = "278",
    pages = "421--443",
    year = "1993"
}

@article{Shibata:2007zzb,
    author = "Shibata, Masaru",
    title = "{Rotating black hole surrounded by self-gravitating torus in the puncture framework}",
    doi = "10.1103/PhysRevD.76.064035",
    journal = "Phys. Rev. D",
    volume = "76",
    pages = "064035",
    year = "2007"
}

@inproceedings{Gourgoulhon:2010ju,
    author = "Gourgoulhon, Eric",
    title = "{An Introduction to the theory of rotating relativistic stars}",
    booktitle = "{CompStar 2010: School and Workshop on Computational Tools for Compact Star Astrophysics}",
    eprint = "1003.5015",
    archivePrefix = "arXiv",
    primaryClass = "gr-qc",
    month = "3",
    year = "2010"
}

@article{Gourgoulhon:2001ec,
    author = "Gourgoulhon, Eric and Grandclement, Philippe and Bonazzola, Silvano",
    title = "{Binary black holes in circular orbits. 1. A Global space-time approach}",
    eprint = "gr-qc/0106015",
    archivePrefix = "arXiv",
    doi = "10.1103/PhysRevD.65.044020",
    journal = "Phys. Rev. D",
    volume = "65",
    pages = "044020",
    year = "2002"
}

@article{Cook:2004kt,
    author = "Cook, Gregory B. and Pfeiffer, Harald P.",
    title = "{Excision boundary conditions for black hole initial data}",
    eprint = "gr-qc/0407078",
    archivePrefix = "arXiv",
    doi = "10.1103/PhysRevD.70.104016",
    journal = "Phys. Rev. D",
    volume = "70",
    pages = "104016",
    year = "2004"
}

@article{Grandclement:2022wif,
    author = "Grandcl{\'e}ment, Philippe and Nicoules, Jordan",
    title = "{Boundary conditions for stationary black holes: Application to Kerr, Mart{\'\i}nez-Troncoso-Zanelli, and hairy black holes}",
    eprint = "2203.09341",
    archivePrefix = "arXiv",
    primaryClass = "gr-qc",
    doi = "10.1103/PhysRevD.105.104011",
    journal = "Phys. Rev. D",
    volume = "105",
    number = "10",
    pages = "104011",
    year = "2022"
}

@article{Arnowitt:1960zzc,
    author = "Arnowitt, R. and Deser, S. and Misner, C. W.",
    title = "{Energy and the Criteria for Radiation in General Relativity}",
    doi = "10.1103/PhysRev.118.1100",
    journal = "Phys. Rev.",
    volume = "118",
    pages = "1100--1104",
    year = "1960"
}

@article{DeWitt:1967yk,
    author = "DeWitt, Bryce S.",
    editor = "Fang, Li-Zhi and Ruffini, R.",
    title = "{Quantum Theory of Gravity. 1. The Canonical Theory}",
    doi = "10.1103/PhysRev.160.1113",
    journal = "Phys. Rev.",
    volume = "160",
    pages = "1113--1148",
    year = "1967"
}

@article{Arnowitt:1962hi,
    author = "Arnowitt, Richard L. and Deser, Stanley and Misner, Charles W.",
    title = "{The Dynamics of general relativity}",
    eprint = "gr-qc/0405109",
    archivePrefix = "arXiv",
    doi = "10.1007/s10714-008-0661-1",
    journal = "Gen. Rel. Grav.",
    volume = "40",
    pages = "1997--2027",
    year = "2008"
}

@article{Regge:1974zd,
    author = "Regge, Tullio and Teitelboim, Claudio",
    title = "{Role of Surface Integrals in the Hamiltonian Formulation of General Relativity}",
    reportNumber = "Print-74-0988 (IAS,PRINCETON)",
    doi = "10.1016/0003-4916(74)90404-7",
    journal = "Annals Phys.",
    volume = "88",
    pages = "286",
    year = "1974"
}

@book{Gourgoulhon:2012ffd,
    author = "Gourgoulhon, Eric",
    title = "{3+1 Formalism in General Relativity}",
    doi = "10.1007/978-3-642-24525-1",
    publisher = "Springer",
    series = "Lecture Notes in Physics",
    year = "2012"
}

@article{Uryu:1999uu,
    author = "Uryu, Koji and Eriguchi, Yoshiharu",
    title = "{A New numerical method for constructing quasiequilibrium sequences of irrotational binary neutron stars in general relativity}",
    eprint = "gr-qc/9908059",
    archivePrefix = "arXiv",
    doi = "10.1103/PhysRevD.61.124023",
    journal = "Phys. Rev. D",
    volume = "61",
    pages = "124023",
    year = "2000"
}

@article{Ansorg:2003br,
    author = "Ansorg, Marcus and Kleinwachter, A. and Meinel, R.",
    title = "{Highly accurate calculation of rotating neutron stars: detailed description of the numerical methods}",
    eprint = "astro-ph/0301173",
    archivePrefix = "arXiv",
    doi = "10.1051/0004-6361:20030618",
    journal = "Astron. Astrophys.",
    volume = "405",
    pages = "711",
    year = "2003"
}

@article{Renkhoff:2023nfw,
    author = {Renkhoff, Sarah and Cors, Daniela and Hilditch, David and Br{\"u}gmann, Bernd},
    title = "{Adaptive hp refinement for spectral elements in numerical relativity}",
    eprint = "2302.00575",
    archivePrefix = "arXiv",
    primaryClass = "gr-qc",
    doi = "10.1103/PhysRevD.107.104043",
    journal = "Phys. Rev. D",
    volume = "107",
    number = "10",
    pages = "104043",
    year = "2023"
}

@article{Ansorg:2005bp,
    author = "Ansorg, Marcus",
    title = "{A Double-domain spectral method for black hole excision data}",
    eprint = "gr-qc/0505059",
    archivePrefix = "arXiv",
    reportNumber = "AEI-2005-102",
    doi = "10.1103/PhysRevD.72.024018",
    journal = "Phys. Rev. D",
    volume = "72",
    pages = "024018",
    year = "2005"
}

@article{Kuan:2023hrh,
    author = "Kuan, Hao-Jui and Van Aelst, Karim and Lam, Alan Tsz-Lok and Shibata, Masaru",
    title = "{Binary neutron star mergers in massive scalar-tensor theory: Quasiequilibrium states and dynamical enhancement of the scalarization}",
    eprint = "2309.01709",
    archivePrefix = "arXiv",
    primaryClass = "gr-qc",
    doi = "10.1103/PhysRevD.108.064057",
    journal = "Phys. Rev. D",
    volume = "108",
    number = "6",
    pages = "064057",
    year = "2023"
}

@article{Kuan:2024jnw,
    author = "Kuan, Hao-Jui and Kiuchi, Kenta and Shibata, Masaru",
    title = "{Tidal Resonance in Binary Neutron Star Inspirals: A High-Precision Study in Numerical Relativity}",
    eprint = "2411.16850",
    archivePrefix = "arXiv",
    primaryClass = "hep-ph",
    doi = "10.1103/j3zk-z17h",
    journal = "Phys. Rev. Lett.",
    volume = "135",
    number = "14",
    pages = "141403",
    year = "2025"
}

@misc{Kuan:2025bzu,
    author = "Kuan, Hao-Jui and Markin, Ivan and Ujevic, Maximiliano and Dietrich, Tim and Kiuchi, Kenta and Shibata, Masaru and Tichy, Wolfgang",
    title = "{The error budget of binary neutron star merger simulations for configurations with high spin}",
    eprint = "2506.02115",
    archivePrefix = "arXiv",
    primaryClass = "gr-qc",
    month = "6",
    year = "2025"
}

@article{Blanchet:2013haa,
    author = "Blanchet, Luc",
    title = "{Post-Newtonian Theory for Gravitational Waves}",
    eprint = "1310.1528",
    archivePrefix = "arXiv",
    primaryClass = "gr-qc",
    doi = "10.12942/lrr-2014-2",
    journal = "Living Rev. Rel.",
    volume = "17",
    pages = "2",
    year = "2014"
}

@article{Kidder:1995zr,
    author = "Kidder, Lawrence E.",
    title = "{Coalescing binary systems of compact objects to postNewtonian 5/2 order. 5. Spin effects}",
    eprint = "gr-qc/9506022",
    archivePrefix = "arXiv",
    reportNumber = "NU-GR-11, WUGRAV-94-6A",
    doi = "10.1103/PhysRevD.52.821",
    journal = "Phys. Rev. D",
    volume = "52",
    pages = "821--847",
    year = "1995"
}

@article{Bohe:2012mr,
    author = "Bohe, Alejandro and Marsat, Sylvain and Faye, Guillaume and Blanchet, Luc",
    title = "{Next-to-next-to-leading order spin-orbit effects in the near-zone metric and precession equations of compact binaries}",
    eprint = "1212.5520",
    archivePrefix = "arXiv",
    primaryClass = "gr-qc",
    doi = "10.1088/0264-9381/30/7/075017",
    journal = "Class. Quant. Grav.",
    volume = "30",
    pages = "075017",
    year = "2013"
}

@article{Pfeiffer:2007yz,
    author = "Pfeiffer, Harald P. and Brown, Duncan A. and Kidder, Lawrence E. and Lindblom, Lee and Lovelace, Geoffrey and Scheel, Mark A.",
    editor = "Campanelli, Manuela and Rezzolla, Luciano",
    title = "{Reducing orbital eccentricity in binary black hole simulations}",
    eprint = "gr-qc/0702106",
    archivePrefix = "arXiv",
    doi = "10.1088/0264-9381/24/12/S06",
    journal = "Class. Quant. Grav.",
    volume = "24",
    pages = "S59--S82",
    year = "2007"
}

@article{Purrer:2012wy,
    author = "Purrer, Michael and Husa, Sascha and Hannam, Mark",
    title = "{An Efficient iterative method to reduce eccentricity in numerical-relativity simulations of compact binary inspiral}",
    eprint = "1203.4258",
    archivePrefix = "arXiv",
    primaryClass = "gr-qc",
    doi = "10.1103/PhysRevD.85.124051",
    journal = "Phys. Rev. D",
    volume = "85",
    pages = "124051",
    year = "2012"
}

@article{Ramos-Buades:2018azo,
    author = "Ramos-Buades, Antoni and Husa, Sascha and Pratten, Geraint",
    title = "{Simple procedures to reduce eccentricity of binary black hole simulations}",
    eprint = "1810.00036",
    archivePrefix = "arXiv",
    primaryClass = "gr-qc",
    doi = "10.1103/PhysRevD.99.023003",
    journal = "Phys. Rev. D",
    volume = "99",
    number = "2",
    pages = "023003",
    year = "2019"
}

@article{Habib:2024soh,
    author = "Habib, Sarah and Scheel, Mark A. and Teukolsky, Saul A.",
    title = "{Eccentricity reduction for quasicircular binary evolutions}",
    eprint = "2410.05531",
    archivePrefix = "arXiv",
    primaryClass = "gr-qc",
    doi = "10.1103/PhysRevD.111.084059",
    journal = "Phys. Rev. D",
    volume = "111",
    number = "8",
    pages = "084059",
    year = "2025"
}

@article{Kyutoku:2014yba,
    author = "Kyutoku, Koutarou and Shibata, Masaru and Taniguchi, Keisuke",
    title = "{Reducing orbital eccentricity in initial data of binary neutron stars}",
    eprint = "1405.6207",
    archivePrefix = "arXiv",
    primaryClass = "gr-qc",
    doi = "10.1103/PhysRevD.90.064006",
    journal = "Phys. Rev. D",
    volume = "90",
    number = "6",
    pages = "064006",
    year = "2014"
}

@article{Kyutoku:2020lgg,
    author = "Kyutoku, Koutarou and Kawaguchi, Kyohei and Kiuchi, Kenta and Shibata, Masaru and Taniguchi, Keisuke",
    title = "{Reducing orbital eccentricity in initial data of black hole{\textendash}neutron star binaries in the puncture framework}",
    eprint = "2009.03896",
    archivePrefix = "arXiv",
    primaryClass = "gr-qc",
    doi = "10.1103/PhysRevD.103.023002",
    journal = "Phys. Rev. D",
    volume = "103",
    number = "2",
    pages = "023002",
    year = "2021"
}

@article{Blanchet:2001id,
    author = "Blanchet, Luc",
    title = "{Innermost circular orbit of binary black holes at the third postNewtonian approximation}",
    eprint = "gr-qc/0112056",
    archivePrefix = "arXiv",
    doi = "10.1103/PhysRevD.65.124009",
    journal = "Phys. Rev. D",
    volume = "65",
    pages = "124009",
    year = "2002"
}

@misc{Han:2026vnm,
    author = "Han, Ming-Zhe and Kiuchi, Kenta and Shibata, Masaru",
    title = "{SACRA-K: A Performance-Portable Numerical Relativity Code with Kokkos}",
    eprint = "2607.08743",
    archivePrefix = "arXiv",
    primaryClass = "astro-ph.HE",
    month = "7",
    year = "2026"
}

@article{Yamamoto:2008js,
    author = "Yamamoto, Tetsuro and Shibata, Masaru and Taniguchi, Keisuke",
    title = "{Simulating coalescing compact binaries by a new code SACRA}",
    eprint = "0806.4007",
    archivePrefix = "arXiv",
    primaryClass = "gr-qc",
    doi = "10.1103/PhysRevD.78.064054",
    journal = "Phys. Rev. D",
    volume = "78",
    pages = "064054",
    year = "2008"
}

@article{Kiuchi:2017pte,
    author = "Kiuchi, Kenta and Kawaguchi, Kyohei and Kyutoku, Koutarou and Sekiguchi, Yuichiro and Shibata, Masaru and Taniguchi, Keisuke",
    title = "{Sub-radian-accuracy gravitational waveforms of coalescing binary neutron stars in numerical relativity}",
    eprint = "1708.08926",
    archivePrefix = "arXiv",
    primaryClass = "astro-ph.HE",
    doi = "10.1103/PhysRevD.96.084060",
    journal = "Phys. Rev. D",
    volume = "96",
    number = "8",
    pages = "084060",
    year = "2017"
}

@article{Kiuchi:2019kzt,
    author = "Kiuchi, Kenta and Kawaguchi, Kyohei and Kyutoku, Koutarou and Sekiguchi, Yuichiro and Shibata, Masaru",
    title = "{Sub-radian-accuracy gravitational waves from coalescing binary neutron stars in numerical relativity. II. Systematic study on the equation of state, binary mass, and mass ratio}",
    eprint = "1907.03790",
    archivePrefix = "arXiv",
    primaryClass = "astro-ph.HE",
    doi = "10.1103/PhysRevD.101.084006",
    journal = "Phys. Rev. D",
    volume = "101",
    number = "8",
    pages = "084006",
    year = "2020"
}

@article{Kiuchi:2022ubj,
    author = "Kiuchi, Kenta and Held, Loren E. and Sekiguchi, Yuichiro and Shibata, Masaru",
    title = "{Implementation of advanced Riemann solvers in a neutrino-radiation magnetohydrodynamics code in numerical relativity and its application to a binary neutron star merger}",
    eprint = "2205.04487",
    archivePrefix = "arXiv",
    primaryClass = "astro-ph.HE",
    doi = "10.1103/PhysRevD.106.124041",
    journal = "Phys. Rev. D",
    volume = "106",
    number = "12",
    pages = "124041",
    year = "2022"
}

@article{Kiuchi:2025ksk,
    author = "Kiuchi, Kenta",
    title = "{Toward high-precision inspiral gravitational waveforms from binary neutron star mergers in numerical relativity}",
    eprint = "2508.10981",
    archivePrefix = "arXiv",
    primaryClass = "astro-ph.HE",
    doi = "10.1103/zmdc-xkcm",
    journal = "Phys. Rev. D",
    volume = "112",
    number = "8",
    pages = "084008",
    year = "2025"
}

@article{Nakano:2015pta,
    author = "Nakano, Hiroyuki and Healy, James and Lousto, Carlos O. and Zlochower, Yosef",
    title = "{Perturbative extraction of gravitational waveforms generated with Numerical Relativity}",
    eprint = "1503.00718",
    archivePrefix = "arXiv",
    primaryClass = "gr-qc",
    doi = "10.1103/PhysRevD.91.104022",
    journal = "Phys. Rev. D",
    volume = "91",
    number = "10",
    pages = "104022",
    year = "2015"
}

@article{Reisswig:2010di,
    author = "Reisswig, Christian and Pollney, Denis",
    title = "{Notes on the integration of numerical relativity waveforms}",
    eprint = "1006.1632",
    archivePrefix = "arXiv",
    primaryClass = "gr-qc",
    doi = "10.1088/0264-9381/28/19/195015",
    journal = "Class. Quant. Grav.",
    volume = "28",
    pages = "195015",
    year = "2011"
}

@article{OShaughnessy:2011pmr,
    author = "O'Shaughnessy, R. and Vaishnav, B. and Healy, J. and Meeks, Z. and Shoemaker, D.",
    title = "{Efficient asymptotic frame selection for binary black hole spacetimes using asymptotic radiation}",
    eprint = "1109.5224",
    archivePrefix = "arXiv",
    primaryClass = "gr-qc",
    reportNumber = "LIGO-DCC-P1100113",
    doi = "10.1103/PhysRevD.84.124002",
    journal = "Phys. Rev. D",
    volume = "84",
    pages = "124002",
    year = "2011"
}

@article{Apostolatos:1994mx,
    author = "Apostolatos, Theocharis A. and Cutler, Curt and Sussman, Gerald J. and Thorne, Kip S.",
    title = "{Spin induced orbital precession and its modulation of the gravitational wave forms from merging binaries}",
    reportNumber = "GRP-382",
    doi = "10.1103/PhysRevD.49.6274",
    journal = "Phys. Rev. D",
    volume = "49",
    pages = "6274--6297",
    year = "1994"
}

@article{Dietrich:2019kaq,
    author = "Dietrich, Tim and Samajdar, Anuradha and Khan, Sebastian and Johnson-McDaniel, Nathan K. and Dudi, Reetika and Tichy, Wolfgang",
    title = "{Improving the NRTidal model for binary neutron star systems}",
    eprint = "1905.06011",
    archivePrefix = "arXiv",
    primaryClass = "gr-qc",
    doi = "10.1103/PhysRevD.100.044003",
    journal = "Phys. Rev. D",
    volume = "100",
    number = "4",
    pages = "044003",
    year = "2019"
}

\end{document}